\documentclass{JHEP3}
\pdfoutput=1
\usepackage{bm}         
\usepackage{epsfig}
\usepackage{amsmath, amsfonts}
\usepackage{epsfig}
\usepackage{cite}

\providecommand{\openone}{\leavevmode\hbox{\large1\kern-7.3pt\normalsize1}}
\newcommand{\be}{\begin{equation}}
\newcommand{\ee}{\end{equation}}
\newcommand{\ba}{\begin{eqnarray}}
\newcommand{\ea}{\end{eqnarray}}

\newcommand{\tr}{{\rm Tr\,}}
\newcommand{\nn}{\nonumber \\}
\newcommand{\fr}[2]{{\frac{#1}{#2}\,}}

\renewcommand{\vec}[1]{{\bf #1}}
\renewcommand{\(}{\left(}
\renewcommand{\)}{\right)}

\newcommand{\lk}{\left[}
\newcommand{\rk}{\right]}

\def\be{\begin{equation}}
\def\ee{\end{equation}}

\def\SOsix{\mbox{SO}(6)}
\def\SOfour{\mbox{SO}(4)}
\def\SUN{\mbox{SU}(N)}
\def\gYM{g_{\rm{YM}}}

\preprint{CERN-PH-TH/2008-022 \\
NSF-KITP-08-28 \\
PUPT-2256 \\
TUW-08-03\\
arXiv.org/0802.2956 [hep-th]}

\title
    {%
    Remarks on Heavy-Light Mesons from AdS/CFT
     }

\author
    {%
    Christopher P.~Herzog$^1$, Stefan A.~Stricker$^2$ and Aleksi Vuorinen$^{23}$    \\
   $^1$Joseph Henry Laboratories, Princeton University, Princeton, NJ 08544, USA \\
$^2$Institut f\"ur Theoretische Physik, Technische Universit\"at Wien, Wiedner Hauptstr.\\
\hskip 0.05in 8-10, A-1040 Vienna, Austria \\
$^3$CERN, Physics Department, TH Unit, CH-1211 Geneva 23, Switzerland\\
    }

\abstract{
We use the AdS/CFT correspondence to compute the energy spectrum of heavy-light mesons
in a ${\mathcal N}=2$ $\SUN$ super Yang-Mills theory with two massive hypermultiplets.
In the heavy quark limit, similar to QCD, we find that the excitation
energies are independent of the heavy quark mass. We also make some remarks about related AdS/CFT
models of flavor with less supersymmetry.
}


\begin{document}

\section{Introduction}

The heavy quark limit of QCD has been an important tool in understanding the spectrum
and decays of mesons and baryons with a heavy quark constituent; see Ref.~\cite{Neubert:1993mb}
for a review.  When the mass of the heavy quark is large compared to the QCD scale,
$m_h \gg \Lambda_{\rm{QCD}}$, the interaction between the heavy quark
and the light quarks and gluons becomes independent of the spin and flavor of the heavy
quark. This independence yields predictions for the $m_h$ dependence of
the meson spectrum and weak decay amplitudes.
In this paper, we investigate the heavy quark limit not in QCD
but in a cousin of ${\mathcal N}=4$ $\SUN$ super Yang-Mills (SYM) theory.
We add two fundamental hypermultiplets, with masses $m_l$ and $m_h$,
to ${\mathcal N}=4$ SYM, breaking the supersymmetry to ${\mathcal N}=2$.
Using the AdS/CFT correspondence \cite{Maldacena:1997re,Witten:1998qj,Gubser:1998bc},
we study the spectrum of heavy-light mesons in this theory at large $N$ and large 't Hooft
coupling $\lambda = \gYM^2 N$.

One reason why heavy quarks are easier to understand in QCD than light quarks is asymptotic
freedom; at short distance scales and high energies, the strong force becomes weak.
Roughly speaking, for energies sufficiently above $\Lambda_{\rm{QCD}}$, the coupling constant
$\alpha_s$ becomes small, and thus the interactions of the heavy quarks, charm, bottom and top, are
governed by a weak effective coupling $\alpha_s(m_h)$. The light quarks,
up, down, and strange, on the other hand experience a much stronger coupling $\alpha_s(\Lambda)$,
with $\Lambda$  only slightly above $\Lambda_{\rm{QCD}}$, where the coupling diverges.
Indeed, the strong force between two heavy quarks is weak enough to be treated
perturbatively, and is similar to the force between an electron and a positron.
Heavy-heavy mesons, which are bound states of two heavy quarks, therefore have
measured properties very similar to positronium.\footnote{%
 Note however that highly excited charmonium and bottomonium states
 are expected to be sensitive to the details of confinement.  For these
 excited states, the quarks are separated by relatively large distances and experience
 a linear confining potential rather than a Coulombic potential.
 To reproduce the full spectrum, the Cornell potential, which interpolates between
 these two limiting forms, is often used.
}

Heavy-light mesons, in contrast, are more complicated objects, as their light quark constituent
experiences strong interactions.  Qualitatively, the heavy quark is a small object
of size $1/m_h$ surrounded by a ``brown muck'' of size $1/\Lambda_{\rm{QCD}}$
of virtual strongly interacting light quarks, antiquarks, and gluons.
However, the small size of the heavy quark leads to simplifications.  The ``brown muck''
cannot resolve the spin or flavor of the heavy quark to leading order in $1/m_h$,
which means the interaction is spin and flavor blind.

The current paper was motivated by wondering, what parallels exist between
heavy-light mesons in real world QCD and in strongly coupled ${\mathcal N}=2$ $\SUN$
SYM theory with two massive hypermultiplets.
The parent theory ${\mathcal N}=4$ $\SUN$ SYM is clearly very different from QCD.
Most importantly for our comparison,
${\mathcal N}=4$ SYM is conformal, and we thus have no equivalent notion of
the coupling constant being $m_h$ dependent.  We also have no notion of
a confinement or QCD scale $\Lambda_{\rm{QCD}}$; for us the IR scale will
be $m_l$. It is true that
adding $N_f = 2$ hypermultiplets to ${\mathcal N}=4$ SYM breaks the conformal
symmetry, but the nonzero beta
function in fact runs in the wrong direction, toward strong coupling in the UV.
In this paper, we will, however, work in the limit $N_f \ll N$, and therefore ignore $N_f/N$ suppressed effects.

Despite these differences, there is persistent hope that we may gain insights into QCD by asking the right questions
about ${\mathcal N}=4$ SYM and its relatives at strong coupling.
For example, at zero temperature, the Klebanov-Strassler model \cite{KS}
provides a geometric understanding of abelian chiral symmetry breaking
and confinement for a ${\mathcal N}=1$ supersymmetric gauge theory in this AdS/CFT context.
Regarding nonzero temperature physics, where the arguments are perhaps
more compelling, Ref.~\cite{Shuryak}
made the following two observations.
First, consider the ratio of the pressure at strong and weak coupling.
The ratio for ${\mathcal N}=4$ SYM
was computed by Ref.~\cite{GubserKlebanovPeet} to be 3/4.
QCD is not conformal, but lattice simulations can be used to compute the
pressure at a few times the deconfinement temperature where the theory
is relatively strongly interacting and the pressure slowly varying.  The ratio
of this pressure to the free result is about 0.8.
The second observation is that
at strong coupling, both ${\mathcal N}=4$ SYM  and QCD are believed
to have very small viscosities (see \textit{e.g.~}Refs.~\cite{PSS:PRL,roma}).

The AdS/CFT correspondence maps ${\mathcal N}=4$ $\SUN$ SYM theory
to type IIB string theory in the curved background $AdS_5 \times S^5$.
We will work in the large $N$ and $\lambda$ limit, where the string
theory becomes classical and can be well approximated by supergravity.
As described by Ref.~\cite{KarchKatz}, a hypermultiplet can be added to
the gauge theory by placing a D7 brane in the dual geometry.
The heavy-light mesons we consider then, according to the duality,
correspond to strings stretching between two parallel D7 branes,
and the energy spectrum consists of the
vibrational and rotational modes of the strings.
Consistent with our large $N$ limit, we will neglect the back reaction of
the D branes on the geometry, as well as the back reaction of the strings on the D branes and
the geometry.

Despite the conformal nature of the theory we consider, we find that the meson spectrum
is, in an appropriate sense, spin and flavor blind
in the heavy quark and strong coupling limit.
The mass $M_{hl}$ of the heavy-light mesons we find has the form
\be
M_{hl} = m_h + m_l \, f\left( \frac{J}{\sqrt{\lambda}}, \frac{Q}{\sqrt{\lambda}}, \frac{n}{\sqrt{\lambda}} \right) + {\mathcal O}\left( \frac{m_l^2}{m_h} \right) \ ,
\ee
where $J$ is the angular momentum of the meson, $Q$ an R-charge, and $n$ a quantum number specifying a
radial excitation.\footnote{Recall that ${\mathcal N}=2$ supersymmetric gauge theories
have a global R-symmetry. Geometrically, $Q$ is an angular momentum in the internal $S^5$.}
We have not introduced a confinement scale and thus $m_l$ takes the place of $\Lambda_{\rm{QCD}}$.

One important aspect of this heavy-light meson spectrum is its $m_h$ independence,
which can be understood in the following way. The excitations (at least in $n$ and $J$)
we find are closely analogous to
the modes of a guitar string, the length of which is proportional to $1/m_l - 1/m_h$.  In the heavy
quark limit, the length of the string becomes independent of $1/m_h$, and hence it is expected that also
the frequencies of the modes become $1/m_h$ independent.

After the appearance of Ref.~\cite{KarchKatz},
there have been many detailed studies of the meson spectrum of the ${\mathcal N}=2$
$\SUN$ SYM theory beginning with Refs.~\cite{Karch:2002xe, Kruczenski:2003be}.
In fact, a nice review \cite{Erdmenger:2007cm} has appeared to which
we point the interested reader for a more complete list of references.
To understand what is new about the current paper,
it is useful to outline the differences of our work from Ref.~\cite{Kruczenski:2003be},
where the authors considered the meson spectrum
for ${\mathcal N}=2$ SYM theory with a single massive hypermultiplet of mass $m$.
They considered two different types of mesons. The first type have a very
small mass $M \sim m / \sqrt{\lambda}$ and spin 0, 1/2, or 1, and
are dual to fluctuations of the D7 brane
embedding.  The second type are dual to U-shaped semiclassical strings with much larger
angular momentum $J$ and mass.  For $J \ll \sqrt{\lambda}$, the mass
obeys Regge scaling $M \sim m \sqrt{J} / \lambda^{1/4}$
while for $J \gg \sqrt{\lambda}$, the
potential is Coulombic $M  = 2m - \mbox{const} / J^2$.
While the behavior of these types of mesons
are qualitatively diffferent, there is expected to be a way in which as we consider mesons
with larger and larger angular momentum, the D7 brane fluctuations in fact morph into semiclassical string
configurations.

The ground state of our heavy-light meson is a string, which stretches between two
D7 branes separated by a finite distance proportional to the mass difference between
the hypermultiplets.  Having taken the heavy-quark limit, there is no sense in which
our meson spectrum is well approximated by D7 brane fluctuations. To find the spectrum,
we therefore instead consider fluctuations of the string itself, which will correspond to radial
excitations of the meson. We also consider the dependence of the string energy on its
angular momentum $J$ and charge $Q$, and this part of the analysis is similar to the
second half of Ref.~\cite{Kruczenski:2003be} and Section 2 of \cite{ParedesTalavera}.

The types of heavy-light mesons we consider have been studied before, in
Refs.~\cite{ Erdmengerheavylight, ErdmengerDBI, ParedesTalavera}.
Ref.~\cite{ParedesTalavera}, is very similar in spirit
to ours. Indeed, Section 2 of Ref.~\cite{ParedesTalavera}
overlaps to some extent with our discussion of the spinning strings in Section \ref{sec:spinning}.1.
In Refs.~\cite{Erdmengerheavylight, ErdmengerDBI},
it was pointed out that the ground state heavy-light mesons
have a mass which scales as the difference of the heavy quark masses,
$M = m_h - m_l$.  This scaling is very different from the
D7 brane fluctuations considered in Ref.~\cite{Kruczenski:2003be}, which
yielded masses $M \sim m / \sqrt{\lambda}$ for the heavy-heavy and light-light
mesons.  Ref.~\cite{Erdmengerheavylight} also demonstrated that
the excitation energies above the ground state are suppressed by a power of $\lambda$.
This work should in principle be very similar to what we do here, as the authors of 
Ref.~\cite{Erdmengerheavylight}
also study the fluctuation spectrum of a semiclassical string stretching between
two D7 branes in the $AdS_5 \times S^5$ geometry. However, they
work in an approximation where the strings do not bend and find that
the excitation energies for heavy-light mesons
scale with $m_h$ instead of $m_l$.
Ref.~\cite{ErdmengerDBI} in contrast is a calculation in a different limit:
They consider the case where the masses of the two
hypermultiplets become degenerate and thus non-abelian effects on the
D7 branes are important.

Our paper is organized as follows. We begin in Section \ref{sec:SUSY} by reviewing the dual supergravity construction of
${\mathcal N}=2$ SYM theory with two massive fundamental hypermultiplets, and in addition we make some remarks
about related constructions that preserve only ${\mathcal N}=1$ supersymmetry, allowing
for a novel way of thinking about meson decay and also yielding a spectrum of heavy-light
mesons similar to the spectrum of the heavy-heavy and light-light mesons considered
 in \cite{Kruczenski:2003be}.
 In the following sections we consider only the ${\mathcal N}=2$ supersymmetry preserving case.
Section \ref{sec:prelims} fixes our notation and sets up the supergravity calculation of
the heavy-light meson spectrum.

In Section \ref{sec:fluctuations}, we analyze small fluctuations of the string dual to the
heavy-light meson and thus obtain the spectrum as a function of what we called $n$
above.
This analysis ignores nonlinearities in the equation of motion for the string and is
valid when the occupation numbers of the modes are small compared to $\sqrt{\lambda}$.
Section \ref{sec:spinning} follows with a discussion of spinning strings dual
to heavy-light mesons with large charge and angular momentum.  The analysis
is purely classical but employs the full nonlinear equations of motion.  We expect
a classical analysis to be valid in the limit where $J \gg 1$ and $Q \gg 1$, but we also find
that the solutions match smoothly onto the small fluctuations considered in
Section \ref{sec:fluctuations} at small values of $J$ and $Q$.
The paper concludes with a
comparison to the spectrum of real world (QCD) heavy-light mesons in the Summary section.

\section{Supersymmetry considerations}
\label{sec:SUSY}

We know that type IIB strings in an $AdS_5 \times S^5$ space-time are dual to
${\mathcal N}=4$ $\SUN$ super Yang-Mills theory through the AdS/CFT correspondence.
The space $AdS_5 \times S^5$ has the line element
\be
ds^2 = L^2 \left[ u^2 \eta_{\mu\nu} dx^\mu dx^\nu + \frac{ \delta_{ij} dy^i dy^j}{u^2} \right] \ ,
\label{warpedproduct}
\ee
where the indices $i$ and $j$ run from one to six, $\mu$ and $\nu$ run from zero to three,
and $L$ is the radius of curvature.  The coordinate $u^2 \equiv \sum_i (y^i)^2$ is a radial coordinate,
and as $u \to \infty$, we reach the boundary of $AdS_5$.
In this notation, the metric is clearly a warped product of Minkowski space ${\mathbb R}^{1,3}$
with ${\mathbb R}^6$.
The line element can also be written to make the $AdS_5$ more explicit:
\be
ds^2 = \frac{L^2}{z^2} ( \eta_{\mu\nu} dx^\mu dx^\nu + dz^2) + L^2 d \Omega^2 \ ,
\ee
where $d\Omega^2$ is a line element on the $S^5$ and $u= 1/z$.  The $\SOsix$ isometry
group of the $S^5$ geometrically realizes the $\SOsix$ R-symmetry of the dual field theory.

As described by Karch and Katz \cite{KarchKatz}, adding an ${\mathcal N}=2$
hypermultiplet to the gauge theory is dual to placing a D7 brane in the
dual geometry.  The D7 brane spans the Minkowski directions $x^\mu$
and four of the remaining directions in ${\mathbb R}^6$.  With this ansatz,
the D7 brane is insensitive to the RR-five form flux in the curved geometry,
and its behavior is determined solely through the DBI action
\be
S_{DBI} = -\tau_7 \int d^8\xi \sqrt{-\mbox{det} (G_{ab} + 2 \pi \alpha' {\mathcal F}_{ab}) } \ ,
\ee
where $\tau_7 = 1 / (2 \pi)^7 \alpha'^4 g_s$
is the D7 brane tension, $1/2 \pi \alpha'$ is the string tension, $g_s$ is the string
coupling constant,
$G_{ab}$ is the induced metric on the
D7 brane, and ${\mathcal F}_{ab}$ is the gauge field on the D7 brane.
We will consider only the case ${\mathcal F}_{ab}=0$ in these remarks.
Recall that the AdS/CFT dictionary relates
\be
\frac{L^2}{\alpha'} = \sqrt{\lambda} \; \; \; \mbox{and} \; \; \; 4 \pi g_s = \gYM^2 \ ,
\ee
where $\lambda = \gYM^2 N$ is the 't Hooft coupling.

To correspond to a hypermultiplet, the D7 brane must span ${\mathbb R}^{1,3}$,
and thus the four remaining
dimensions of the D7 brane lie in ${\mathbb R}^6$.  It seems natural to choose a gauge in which
four of the coordinates on the D7 brane are the $x^\mu$.  Moreover we pick an embedding in
${\mathbb R}^6$ that does not depend on the $x^\mu$.  Given this independence, the determinant
of the induced metric on the D7 brane will not depend on the warp factor $u^2$ in the
ten dimensional metric (\ref{warpedproduct}).  Dividing out by the volume of Minkowski space,
the DBI action can be written in the form
\be
S_{DBI} = -\tau_7 L^8 \int d^4\xi  \sqrt{  \mbox{det} \left( \frac{\partial \vec y}{\partial \xi^a}{\cdot}   \frac{\partial \vec y}{\partial \xi^b} \right) } \ .
\ee
The D7 brane will satisfy the same
equations of motion that it does in flat space;  the D7 brane describes
 a minimal four dimensional hypersurface in ${\mathbb R}^6$.  Note that the normalization of the
 DBI action can be written in gauge theory language as
 \[
 \tau_7 L^8 =\frac{2\lambda N}{(2 \pi)^6} \ .
 \]
The DBI action is smaller by a factor of $N$ compared to the supergravity action, justifying
our neglect of the back reaction of the D7 brane on the geometry.

 A particularly simple class of hypersurfaces which satisfy the equations of motion
 are surfaces described by a holomorphic embedding equation.
 If we think of ${\mathbb R}^6 = {\mathbb C}^3$ as a complex manifold
and define coordinates $w^j = y^{2j-1} + i y^{2j}$, a D7 brane which
satisfies an equation of the form $f(w^1, w^2, w^3)=0$ for an arbitrary
function $f$ will locally satisfy the equations of motion away from singularities.

The Karch-Katz D7 brane is a hyperplane
described by two linear equations
 $\vec a_1 \cdot \vec y = c_1$ and $\vec a_2 \cdot \vec y = c_2$.
Given the $\SOsix$ rotational symmetry of the sphere, such a hyperplane
can be rotated so that the two equations become $y^5 = c$ and $y^6 = 0$.\footnote{%
Use the $\SOsix$ symmetry to rotate $\vec a_1$ into the $y^5$ direction and
$\vec a_2$ into the $y^5$--$y^6$ plane.  The problem reduces to considering
the intersection of two lines in a plane.  There is a residual $SO(2)$ symmetry
in the $y^5$--$y^6$ plane which always allows us to rotate the intersection point  onto
the $y^5$ axis.
}
In complex coordinates, the hyperplane is
the complex submanifold described by $f = w^3 - c$.  The parameter
$c$ is dual to the mass of the hypermultiplet.

The Karch-Katz D7 brane preserves ${\mathcal N}=2$ supersymmetry, while
the more general case $f(w^1, w^2, w^3)=0$ preserves only ${\mathcal N}=1$
supersymmetry (see \textit{e.g.~}Ref.~\cite{Gomis:2005wc}).  In brief, there are
32 real spinors generating supersymmetry transformations that leave
invariant
the $AdS_5\times S^5$ type IIB supergravity background, 16 of which correspond to ordinary
supercharges and the remainder of which are superconformal.
This number of supercharges is sufficient to generate the
 ${\mathcal N}=4$ superconformal algebra of the dual
Yang-Mills field theory.  Of these 32 spinors, only four of the ordinary
and none of the superconformal generate
supersymmetry transformations which leave
a general D7 brane satisfying $f(w^1, w^2, w^3)=0$
invariant.  The four invariant
spinors are independent of the choice of $f(w^1, w^2, w^3)$.  The Karch-Katz
D7 brane, on the other hand, is left invariant by 8 of the ordinary spinors.

Given that a single Karch-Katz D7 brane corresponds to adding a single
${\mathcal N}=2$ hypermultiplet, adding two such D7 branes should
correspond to adding two hypermultiplets.  In the literature
\cite{Erdmengerheavylight, ErdmengerDBI,
Herzog:2006gh},
we find that the second D7 brane is usually added in a way
such that the embedding equation for the second D7 brane is
parallel to the first, $w^3 = c'$ where $c' \in {\mathbb R}$.
Adding the second D7 brane in such a way has a number of desirable
features.  The theory remains ${\mathcal N}=2$ supersymmetric.
Moreover, an unbroken $\SOfour \subset \SOsix$ of the global
R-symmetry is preserved.
Note that
$c' \in {\mathbb C}$ still preserves ${\mathcal N}=2$ supersymmetry
and the $\SOfour$ R-symmetry.  The relative phase of $c$ and $c'$
affects the relative phase of the hypermultiplet masses and also
the mass of the heavy-light meson, a fact we will return to in the
discussion.

However, a generic second D7 brane would not  be parallel to the first.
Assuming the second D7 brane is also described by a four dimensional
hyperplane inside ${\mathbb R}^6$, the two D7 branes will generically
intersect along a plane ${\mathbb R}^2$.
Such an intersection generically breaks all the supersymmetry.
If supersymmetry is broken, then there will probably be a tachyon, \textit{i.e.~}an instability,
and the
D7 branes will recombine; it's not clear what the final state will be,
and we have little to say about this nonsupersymmetric situation.

While the remaining $\SOfour$ symmetry
is not enough to guarantee
the second Karch-Katz D7 brane can be described by a complex equation as well,
there will be a special case where both D7 brane embeddings
are described by complex equations in ${\mathbb C}^3$.  This special case preserves
${\mathcal N}=1$ supersymmetry.
Indeed, if we add any number of Karch-Katz D7 branes such that they are all described
by complex equations in ${\mathbb C}^3$, ${\mathcal N}=1$ supersymmetry is preserved.
The reason is that the four spinors preserved by both the supergravity background
and the D7 brane are independent of the choice of $f(w^1, w^2, w^3)$.
These intersecting brane configurations
should lead to a heavy-light meson spectrum similar to the heavy-heavy and light-light
meson spectra found in Ref.~\cite{Kruczenski:2003be}.  There will be short strings
localized at the intersection of the two D branes whose masses should scale
as the distance of the intersection from the origin of the geometry divided by
$\sqrt{\lambda}$.  These intersecting configurations also
provide a novel way of thinking about meson
decay, which is different from what has been considered in the literature before
\cite{Cotrone:2005fr,Peeters:2005fq}.
The case of three intersecting Karch-Katz D7 branes would be especially interesting to consider because the
intersection of three four dimensional hyperplanes in ${\mathbb R}^6$ is in general a point.
We, however, leave a study of such spectra and
decays for the future.

Finally, we make a short remark on the field theory aspects of the system we are studying.
We know that ${\mathcal N}=4$ $\SUN$ SYM has the superpotential
\[
W = \tr X[Y,Z]
\]
where $X$, $Y$, and $Z$ are chiral superfields transforming in the adjoint
of $\SUN$.  The Karch-Katz D7 brane leads to the modified
superpotential
\[
W = \tr X[Y,Z] + \tilde Q (m-X)  Q \ ,
\]
where $Q$ and $\tilde Q$ are chiral superfields that transform in the fundamental
of $\SUN$ and combine to form a hypermultiplet.\footnote{%
We have been careless of the
relative normalizations of the different terms in $W$, but they will be fixed by
supersymmetry. See \textit{e.g.~}Ref.~\cite{cv} for details.

}
The ${\mathcal N}=2$ supersymmetry preserving
case of two parallel D7 branes has the superpotential
\[
W = \tr X[Y,Z] +  \tilde Q_h (m_h-X) Q_h  + \tilde Q_l (m_l-X)  Q_l\ .
\]
When $m_h$ and $m_l$ are both real, we chose above
both $c$ and $c' \in {\mathbb R}$.  However, we may introduce a relative
phase between $m_h$ and $m_l$ as well corresponding to $c' \in {\mathbb C}$.
Adding the D7 branes in a way that preserves only
${\mathcal N}=1$ superysmmetry corresponds to more general types of
superpotentials, for example
\[
W = \tr X[Y,Z] +  \tilde Q_h(m_h- X) Q_h  + \tilde Q_l (m_l-Y) Q_l\ .
\]
In most of the rest of what follows, we will restrict to the case where $m_h$ and $m_l$
are real and the two D7 branes preserve ${\mathcal N}=2$ supersymmetry.



\section{Mass spectra of heavy-light mesons: Preliminaries}
\label{sec:prelims}

We consider the special configuration of two parallel D7 branes
in the ${\mathcal N}=2$ supersymmetric scenario described
above where the ground state string will have a nonzero length.
The string hangs from one brane to the other and the string endpoints
correspond to one heavy and one light quark.
Our aim is to derive the mass spectrum of heavy-light mesons by
investigating the spectrum of fluctuations of strings hanging between the
branes.

The $AdS_5\times S_5$
metric (\ref{warpedproduct}) can be thought of as a warped product metric
on ${\mathbb R}^{1,3} \times {\mathbb R}^6$.
We will write the line element on ${\mathbb R}^6$ as
\be
\delta_{ij} dy^i dy^j = d\rho^2 + \rho^2 d\theta^2 + \rho^2 \sin^2 \theta \, d\Omega_2^2 + dy^2 + (dy^6)^2 \ ,
\ee
where $d\Omega_2^2$ is a metric on a unit $S^2$ and we have defined
$\rho^2 \equiv u^2 - (y^5)^2-(y^6)^2$ and $y\equiv y^5$.
The metric on Minkowski space ${\mathbb R}^{1,3}$ we will write as
\be
\eta_{\mu\nu} dx^\mu dx^\nu = -dt^2 + dr^2 + r^2 d\phi^2 + dx^2 \ .
\ee

\FIGURE[t]{
\label{fig:setup}
{\centerline{\def\epsfsize#1#2{0.3#1}
\epsfbox{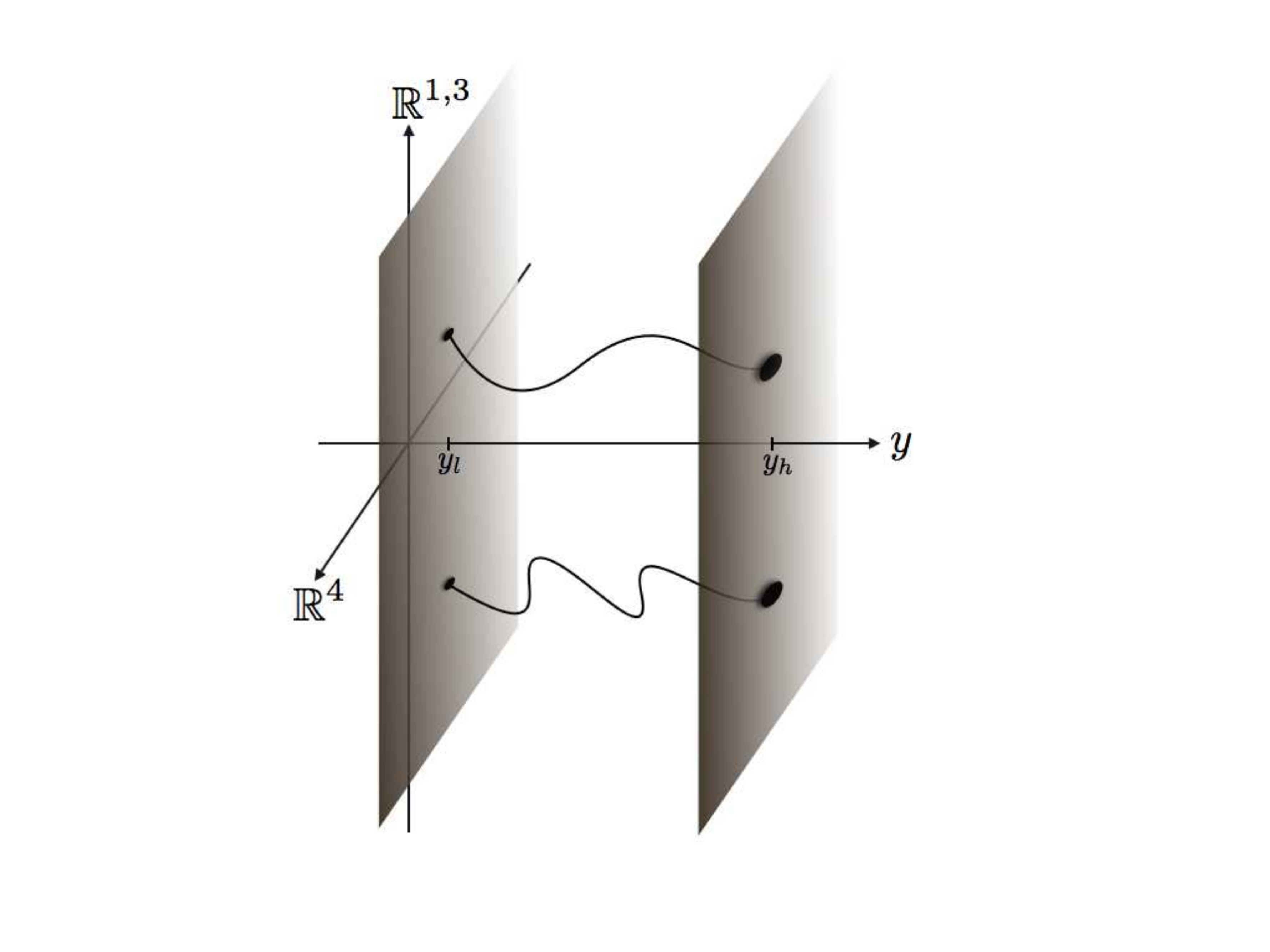}
        }
\caption[a] {A cartoon of our heavy-light mesons as strings stretched between two D7 branes.} }
}

In this geometry, strings that stretch from one D7 brane to another are dual to
mesons, as illustrated in Fig.~\ref{fig:setup} which displays our geometric picture of heavy-light mesons.
Classical strings are described by the Nambu-Goto action
\be
\label{nambugoto}
S_{NG} =
\int d\tau d\sigma {\mathcal L} =
-\frac{1}{2\pi\alpha'} \int d\tau d\sigma \, \sqrt{(\dot X \cdot X')^2 - (\dot X)^2 (X')^2} \ ,
\ee
where $X^A(\tau,\sigma)$ describes the embedding of the string in $AdS_5 \times S^5$.
In our notation, $X \cdot Y = g_{AB} X^A Y^B$ is contracted with the ten dimensional metric, and we
have defined $\partial_\sigma X \equiv X'$ and $\partial_\tau X \equiv \dot X$.
We choose a gauge
in which the worldsheet coordinates are $\tau=t$, $\sigma=y$. The locations of the light and heavy D7 branes will be denoted by $y=y_l$ and $y=y_h$,
and the light and heavy quark masses
\cite{KarchKatz} read
\be
\label{quarkmasses}
m_l = \frac{L^2}{2\pi \alpha'}\; y_l \; ; \;\;\;
m_h = \frac{L^2} {2 \pi \alpha'}\; y_h \; ,\
\ee
where $L^2/ \alpha' = \sqrt{\lambda}$.  The Nambu-Goto action is suppressed by a relative
power of $N$ with respect to the DBI action, and thus we are justified in neglecting
the back reaction of the string on the D7 brane and the geometry in the large $N$ limit.

We wish to study the profile that a string stretching between the D7 branes takes,
assuming that the string sits at a constant position in the internal
unit $S^2$.
The Nambu-Goto action (\ref{nambugoto}) produces the equation of motion
\be
0=\frac{\partial}{\partial \tau} \(g_{AB} \frac{( \dot X \cdot X') (X')^B - (X')^2 \dot X^B}
{  \sqrt{  (\dot X \cdot X')^2 -(\dot X)^2 (X')^2}}\) +
\frac{\partial}{\partial \sigma} \(g_{AB} \frac{( \dot X \cdot X') \dot X^B - (\dot X)^2 (X')^B}
{  \sqrt{ (\dot X \cdot X')^2 -(\dot X)^2 (X')^2}}\) \ ,
\ee
where the various scalar products have the forms
\ba \label{xdotxprime}
\dot{X}\cdot X'&=&L^2\bigg\{u^2 \( \dot{x}x'+ \dot r r' + r^2 \dot \phi \phi' \) +
\fr{1}{u^2}\(\dot{\rho}\rho'+\rho^2\dot\theta \theta'+\dot{y}_6 y'_6\)\bigg\},\\
(\dot{X})^2&=&L^2\bigg\{u^2(-1+\dot{x}^2 + \dot r^2 + r^2 \dot \phi^2)+
\fr{1}{u^2}\(\dot{\rho}^2+\rho^2\dot\theta^2+\dot{y}_6^2\)\bigg\},\\
(X')^2&=&L^2\bigg\{u^2 \( (x')^2+ (r')^2 + r^2 (\phi')^2 \)+
\fr{1}{u^2}\(1+(\rho')^2+\rho^2(\theta')^2+(y'_6)^2\)\bigg\}
\label{xprimesquared}
\ea
and we have rewritten $y^6$ as $y_6$ to avoid confusing superscripts.
The energy and momentum densities of the string are
\be
\pi_A^0 = \frac{\partial {\mathcal L}}{\partial \dot X^A}= -\frac{1}{2\pi\alpha'} g_{AB} \frac{(\dot X \cdot X') (X')^B - (X')^2 (\dot X)^B}{\sqrt{(\dot X \cdot X')^2 - (X')^2 (\dot X)^2}} \ ,
\ee
while the energy and momentum currents read
\be
\pi_A^1 = \frac{\partial {\mathcal L}}{\partial  (X')^A}= -\frac{1}{2\pi\alpha'} g_{AB} \frac{(\dot X \cdot X') (\dot X)^B - (\dot X)^2 (X')^B}{\sqrt{(\dot X \cdot X')^2 - (X')^2(\dot X)^2 }} \ .
\ee
We will apply Neumann boundary conditions in the D7 brane directions at $y=y_l$ and $y=y_h$
\be
\left.  \pi^1_A \right|_{y=y_h,y_l} = 0 \ ,
\ee
for $A = x$, $r$, $\phi$, $\rho$, and $\theta$, implying that
no momentum is assumed to flow into
the string from the D7 brane in these directions.
The coordinate $y_6$ is in contrast subject to Dirichlet
boundary conditions.


\section{Fluctuations in $x$, $\rho$ and $y_6$}
\label{sec:fluctuations}

In this Section, we study radial excitations of the heavy-light mesons.
Specializing to the background of $\dot \theta =0$, $r=0$ and a constant $\rho=\rho_0$, we
consider infinitesimal fluctuations of the string action in the form of
$x=x(t,y)$, $\rho(t,y)=\rho_0+\delta\rho(t,y)$ and $y_6=y_6(t,y)$.
Applying Eqs.~(\ref{xdotxprime})--(\ref{xprimesquared})
where now $u^2 = y^2+(\rho_0+\delta\rho)^2$, we expand the action to second order in the fluctuations, and obtain
\ba
S_{NG} &=& \frac{L^2}{2\pi\alpha'} \int d\tau d\sigma \bigg\{1-\fr{1}{2}\dot{x}^2+\fr{1}{2}u_0^4(x')^2+\fr{1}{2}((\delta\rho')^2+(y'_6)^2)
\nn
&-&\fr{1}{2u_0^4}\(\delta\dot{\rho}^2+\dot{y}_6^2\)\bigg\} \ ,
\ea
with $u_0^2 \equiv y^2+\rho_0^2$.

Translational symmetry in the Minkowski directions guarantees that a constant value of $x$
is a solution to the equation of motion and thus that there is a zero mode in the spectrum
corresponding to motion of the string at constant velocity in the $x$ direction.
Perhaps surprisingly, a constant value of $\rho$ is also a solution and thus there is another
zero mode in the spectrum corresponding to translations of the $\rho$ coordinate, even
though we do not have translational symmetry in these directions.  However, we will see
below that this zero mode is only present for the ground state string.  The fluctuating string
can minimize its energy by moving to $\rho=0$.


There exists an interesting relationship between the equation of motion for the fluctuations in
the $y_6$ and $\delta \rho$ directions and the equation of motion for the fluctuations in $x$
which we believe may be a consequence of supersymmetry.
We will assume that the fluctuations have the time dependence $X^A \sim e^{-i \omega t}$
so that $\ddot X^A = - \omega^2 X^A$.  The equations of motion thus become
\ba
\fr{\partial}{\partial y}\!\!\Big( f(y) \, x^\prime\Big) &=& -\omega^2 x \ , \label{eomone} \\
f(y) \delta \rho'' &=& -\omega^2 \delta  \rho \ ,
\qquad \mbox{and} \qquad
f(y) y_6'' = -\omega^2 y_6 \ ,
\label{eomtwo}
\ea
where $f(y) = (y^2 + \rho_0^2)^2$.
From these expressions, it is clear that if we have a solution $x$ to Eq.~(\ref{eomone}),
then $\delta \rho = f(y) x'$ (or $y_6 = f(y) x'$) satisfies Eq.~(\ref{eomtwo}).  Moreover, given a solution
$\delta \rho$ (or $y_6$) to Eq.~(\ref{eomtwo}), then $x = \delta \rho'$
(or $x = y_6'$) satisfies Eq.~(\ref{eomone}).

A consideration of boundary conditions now reveals that the fluctuations
in $x$ and $y_6$ have the same spectrum up to a zero mode.  While $x$ and $\delta \rho$
satisfy Neumann boundary conditions, $y_6$ satisfies Dirichlet boundary
conditions.  If we solve Eq.~(\ref{eomone}) for the allowed fluctuation modes
$x$ satisfying Neumann boundary conditions,
then the relations between the two equations of motion give us all the
fluctuation modes $y_6$ satisfying Dirichlet boundary conditions.
We have to perform a separate calculation
for the $\delta \rho$ fluctuations, but had the $x$ fluctuations satisfied Dirichlet
boundary conditions instead of Neumann, they would, too, be trivially related
to the $\delta \rho$ fluctuations. We begin with the $x$ fluctuations.


\subsection{The $x$ fluctuations}

The equation (\ref{eomone}) for the $x$ fluctuations can be solved to yield
\ba
x(t,y) &=&\fr{C \rho_0}{\sqrt{y^2+\rho_0^2}}\Bigg\{\sqrt{1+\fr{\omega^2}{\rho_0^2}}
\cos\Bigg[\sqrt{1+\fr{\omega^2}{\rho_0^2}}\arctan\!\bigg[\fr{y}{\rho_0}\bigg]+\alpha\Bigg]\nn
&+&\fr{y}{\rho_0}\sin\Bigg[\sqrt{1+\fr{\omega^2}{\rho_0^2}}\arctan\!\bigg[\fr{y}{\rho_0}\bigg]+\alpha\Bigg]\Bigg\} e^{-i \omega t} \ ,
\ea
where $C$ and $\alpha$ are the two integration constants.
We now apply Neumann boundary conditions $x'(y_l) = x'(y_h)=0$ to determine the allowed
spectrum $\omega$.
Doing this at the light D7 brane, we find
\ba
\label{alphavalue}
\alpha&=&-\sqrt{1+\fr{\omega^2}{\rho_0^2}}\arctan\lk\fr{y_l}{\rho_0}\rk ,
\ea
while applying the boundary conditions at the heavy brane then yields the discrete spectrum:
\be
\omega^x_n=\rho_0\,\sqrt{\fr{n^2\pi^2}{\(\arctan[\rho_0/y_l]-\arctan[\rho_0/y_h]\)^2}-1}\label{omegan}\; ,
\ee
where $n\in{\mathbb Z}^+$. In addition to these values of $n$ however,
the spectrum also contains a zero mode, the trivial solution of $\omega=0$.

Before moving onto the $y_6$ fluctuations, we note that in the $\rho_0= 0$ limit, the mode functions
and spectrum become simpler:
\ba
\label{xfluctzero}
x &=& C (\omega z \cos(\omega(z-z_l)) - \sin (\omega (z-z_l))) e^{-i\omega t} \ , \\
\omega_n^x &=& \frac{\pi n}{z_l-z_h}\ , \qquad \mbox{where} \qquad z=1/y \ .
\ea
The frequencies are the same as those of a guitar string of length $z_l - z_h$, and we thus see that
in the heavy quark limit, $z_h \to 0$, the frequencies become $m_h$ independent.


\subsection{The $y_6$ fluctuations}

The solution to the equation of motion (\ref{eomtwo}) is now related in a trivial way to the
$x$ fluctuations studied above:
\be
y_6 = (y^2+\rho_0^2)^2 x' =
-C \omega^2 \,\sqrt{\rho_0^2+y^2}\sin\Bigg[\sqrt{1+\fr{\omega^2}{\rho_0^2}}\arctan\!\bigg[\fr{y}{\rho_0}\bigg]+\alpha\Bigg] e^{-i \omega t} \ .\label{gsoln}
\ee
In the $\rho_0=0$ limit, the mode function again takes a simpler form
\be
y_6 = \frac{ C \omega^2}{z} \sin (\omega (z-z_l)) e^{-i \omega t} \  \qquad \mbox{where} \qquad z = 1/y \ .
\ee

The Dirichlet boundary conditions $y_6(y_l)=0=y_6(y_h)$ are equivalent to the Neumann
boundary conditions applied to the $x$ fluctuations above, leading to the same
value of $\alpha$ given in Eq.~(\ref{alphavalue}) and the same spectrum
\ba
\omega^{y}_n&=&\rho_0\,\sqrt{\fr{n^2\pi^2}{\(\arctan[\rho_0/y_l]-\arctan[\rho_0/y_h]\)^2}-1} \ ,
\ea
where $n\in{\mathbb Z}^+$. This time, however, there is no zero mode.


\subsection{The $\delta\rho$ fluctuations}
\label{sec:rhofluct}

For the $\delta \rho$ fluctuations, we will not be able to find an analytic spectrum,
but will eventually attempt to understand the spectrum's features both qualitatively and numerically.
We begin with the general solution to Eq.~(\ref{eomtwo}),
\be
\delta \rho(t,y) =C\,\sqrt{\rho_0^2+y^2}\sin\Bigg[\sqrt{1+\fr{\omega^2}{\rho_0^2}}\arctan\!\bigg[\fr{y}{\rho_0}\bigg]+\alpha\Bigg] e^{-i \omega t}\ .
 \label{deltarhosol}
\ee
Applying Neumann boundary conditions at the light brane $\delta\rho'(y_l) = 0$,
we find
\ba
\alpha&=&-\sqrt{1+\fr{\omega^2}{\rho_0^2}}\arctan\!\bigg[\fr{y_l}{\rho_0}\!\bigg]-\arctan\!\Bigg[\sqrt{1+\fr{\omega^2}{\rho_0^2}}\fr{\rho_0}{y_l}\Bigg] \ ,
\ea
while demanding that the boundary conditions are satisfied at the heavy brane leads to
\ba
\label{rhomodes}
&&\tan\Bigg[\sqrt{1+\fr{\omega^2}{\rho_0^2}}\(\arctan\!\bigg[\fr{y_l}{\rho_0}\bigg]-\arctan\!\bigg[\fr{y_h}{\rho_0}\!\bigg]\)+\arctan\!\Bigg[\sqrt{1+\fr{\omega^2}{\rho_0^2}}\fr{\rho_0}{y_l}\Bigg]
\Bigg]\nn
&=&\sqrt{1+\fr{\omega^2}{\rho_0^2}}\fr{\rho_0}{y_h}\ .
\ea
The solutions of this equation give us the spectrum of the fluctuations $\omega^\rho_n$.

Unfortunately, the transcendental nature of the above equation prevents us from solving it analytically. There are, however, various limits, where we can simplify the numerical solution. The first simplification occurs in the limit of large $y_h$, in which one may attempt
a power expansion in $y_l/y_h$. To this end, we write
\ba
\omega^\rho_n&\equiv&\omega_n\,=\, y_l \times \sum_{i=0}^{\infty}\omega_{n,i} \(\fr{y_l}{y_h}\)^i \ , \label{omegaexp}
\ea
substitute this into Eq.~(\ref{rhomodes}), and proceed to solve the equation order by order in the small parameter $y_l/y_h$. At leading order, we easily obtain for $\omega_{n,0}$
\ba
\sqrt{1+\fr{\omega_{n,0}^2\, y_l^2}{\rho_0^2}}\(\fr{\pi}{2}-{\rm arccot} \bigg[\fr{\rho_0}{y_l}\bigg]\)-\arctan\bigg[\sqrt{1+\fr{\omega_{n,0}^2\, y_l^2}{\rho_0^2}}\fr{\rho_0}{y_l}\bigg]&=&n\pi, \label{omega0eq}
\ea
with $n\in \mathbb{Z}^+$. The numerical solution to this equation quickly leads to the forms of the functions $\omega_{n,0}\(\rho_0/y_l\)$.
The next two terms in the power series expansion of Eq.~(\ref{rhomodes}) are solved trivially
by setting $\omega_{n,1}$ and $\omega_{n,2}$ equal to zero, and
it is only at order $i=3$ that we find the next nonzero term in the expansion of Eq.~(\ref{omegaexp}). The forms of the resulting functions $\omega_{n,0}\(\rho_0/y_l\)$ and $\omega_{n,3}\(\rho_0/y_l\)$ will be displayed for $n=1,2,...,5$ in the next Section in a slightly different notation.

One limit, where the functions $\omega_{n,i}$ are in fact analytically solvable is that of large
$\rho_0/y_l$. There, it is straightforward to see that Eq.~(\ref{omega0eq}) reduces to the solution
\ba
\label{largerhoomega}
\omega_{n,0} &=& \sqrt{(2n+1)^2-1}\;\fr{\rho_0}{y_l},
\ea
while the three next orders produce
\be
\omega_{n,1}=\omega_{n,2}\, = \, 0 \qquad \mbox{and} \qquad
\omega_{n,3}= \fr{4}{3\pi}\sqrt{n(n+1)}(2n+1)^2\,\(\fr{\rho_0}{y_l}\)^4 \ .
\ee
It is interesting to contrast Eq.~(\ref{largerhoomega}) with the spectra of the $x$ and $y_6$ fluctuations, which in the same limit ($y_h\rightarrow\infty$ and $\rho_0/y_l$ large) produce from Eq.~(\ref{omegan})
\ba
\omega_n^x &=& \sqrt{(2n)^2-1}\; \rho_0.
\ea
We thus see that at least in this limit, the fluctuation energies in the $x$ and $y_6$ direction lie exactly
in between the energies of the $\rho$ fluctuations.

Finally, we note that in the limit $\rho_0=0$, Eq.~(\ref{deltarhosol}) reduces to
\be
\delta \rho = \frac{-C}{z \sqrt{1+\omega^2 z_l^2 }} \left( \omega z_l \cos (\omega(z-z_l)) + \sin(\omega(z-z_l)) \right) e^{-i \omega t} \ ,
\ee
while condition (\ref{rhomodes}) on the frequencies reduces to the simple expression
\be
\omega (z_h - z_l) = \arctan (\omega z_h) - \arctan(\omega z_l) - \pi n \ , \label{rhozero}
\ee
where $n$ is an integer. This equation, however, is not of an analytically solvable type either, so it must be dealt with numerically.  In the limit $y_h \to \infty$, the first few solutions are
$\omega z_l = 4.493$, 7.725, and 10.904.

\subsection{The meson mass spectrum}

Let us finally look at the energy spectrum of the string fluctuations in more detail.
Using the result
\be
\label{hamiltonian}
E=-\int {\rm d}\sigma \, \pi_t^0 \ ,
\ee
we see that to quadratic order in the fluctuations
the energy of the string can be obtained by integrating the canonical momentum density
\ba
\pi_t^0\,=\,-\fr{L^2}{2\pi \alpha'}\(1+\fr{1}{2}u^4(x')^2+\fr{1}{2}\dot{x}^2+\fr{1}{2}(\delta\rho')^2+\fr{1}{2u^4}(\delta\dot{\rho})^2+\fr{1}{2}(y'_6)^2+\fr{1}{2u^4}(\dot{y}_6)^2\)\!.\nonumber
\ea
From a classical perspective,
the energies will depend on the amplitudes of the fluctuations, while
from a quantum perspective, these amplitudes can only take on discrete values
corresponding to the occupation number of a given mode.
At quadratic order, we essentially have a version of the quantum harmonic oscillator.
The equal time commutation relation $[X^A(y), \pi_A^0(y')] = i \delta(y-y')$ implies, in
units where $\hbar=1$,
that the smallest quanta of excitation are the frequencies we determined before,
the $\omega_n^w$ where $w = x$, $\rho$, or $y$.
We find the simple result
\ba
E &=&m_h-m_l+\sum_{w,n} N_w^n \omega^w_n,
\label{harmresult}
\ea
where $N_w^n$ is the occupation number of the mode $(w,n)$.\footnote{%
 Calculating the zero point energy contribution to these oscillators requires also
 investigating the fermionic fluctuations of the superstring.  We suspect supersymmetry
 implies that the zero point energy vanishes.
}
We therefore note that in order to inspect the mass spectrum of the heavy-light mesons below, we merely need to consider the frequencies $\omega^w_n$ obtained above.
We anticipate Eq.~(\ref{harmresult}) remains valid provided $N_w^n \ll \sqrt{\lambda}$
and we can neglect the nonlinearities in the string equation of motion.


\subsubsection*{The $x$ and $y_6$ fluctuations}
Denoting $q \equiv \rho_0 L^2 / 2 \pi \alpha'$ and using the relation $L^2/\alpha' = \sqrt{\lambda}$, we can write the energy spectrum of the 
$x$ or $y_6$ fluctuations in the
form
\ba
E^x_n= E^y_n &=&m_h-m_l+\fr{2\pi q}{\sqrt{\lambda}}\,\sqrt{\fr{n^2\pi^2}{\(\arctan [q/m_l]-\arctan [q/m_h]\)^2}-1} \ .
\ea
This formula gives the energy
for a string with a single quantum of excitation in the $n$th mode of the $y_6$
or $x$ fluctuations.
In the Introduction, we claimed that in the heavy quark limit, $m_h \gg m_l$,
the energy of the excitations scaled with $m_l$.  Here, seemingly in contradiction
with the earlier claim, we find that in the limit $m_h\gg q$,
we may expand the $\omega^x_n$ in inverse powers of $m_h$, producing
\ba
E^x_n \;=\; m_h-m_l + \fr{2\pi q}{\sqrt{\lambda}} f_n\left(\fr{q}{m_l}\right)+
\fr{2\pi^3 n^2 q^2 }{ \sqrt{\lambda}\, m_h}
\fr{1}{\arctan^3[q/m_l] f_n(q/m_l)}+{\mathcal O}\left(\fr{1}{m_h^2} \right),
\ea
where we have denoted
\ba
f_n(x)&\equiv& \sqrt{\fr{n^2\pi^2}{\arctan^2[x]}-1} \ .
\ea
Thus, the excitation spectrum depends on both light scales $m_l$ and $q$.

We now give two reasons why the scale $q$ should disappear.
First, the derivative of the excitation energies with respect to $q$ is non-negative
\be
\frac{\partial E_n^x}{\partial q} = \frac{\partial E_n^y}{\partial q} \geq 0 \ ,
\ee
and is equal to zero at $q=0$, implying that fluctuations about $q \neq 0$ have
more energy than the equivalent fluctuations about $q=0$.
This inequality suggests that a string fluctuating about a nonzero
value $\rho_0$ will in addition begin to oscillate about $\rho=0$.
In the case of $q=0$, the energy spectra reduce to
\ba
E^x_n = E^y_n &=&m_h-m_l+\fr{m_h m_l}{m_h-m_l}\fr{2\pi^2 n}{\sqrt{\lambda}}\ ,
\ea
where $n \in {\mathbb Z}^+$.  In the heavy quark limit $m_h \gg m_l$, the excitation
spectrum does indeed depend only on $m_l$ to leading order in $m_l/m_h$.

The second reason for the disappearance of the scale $q$ will be developed more in
Section \ref{sec:spinning},
where we will see that for slowly spinning strings in the $\rho$--$\theta$ plane,
a nonzero value of $\rho_0$ is stabilized.  Thus
what would seem to be a zero mode in the $\rho$ direction is lifted and
a continuous change of $q$ will not be possible
for these spinning strings.  However, the stable value of $\rho_0$ is of order
$m_l$ or zero, regardless of the angular momentum, and thus the extra scale $q$
again disappears from the excitation spectrum.

\subsubsection*{The $\delta\rho$ fluctuations}

\FIGURE[t]{
\label{fig:rhofluct}
{ \centerline{\def\epsfsize#1#2{0.7#1}
    \epsfbox{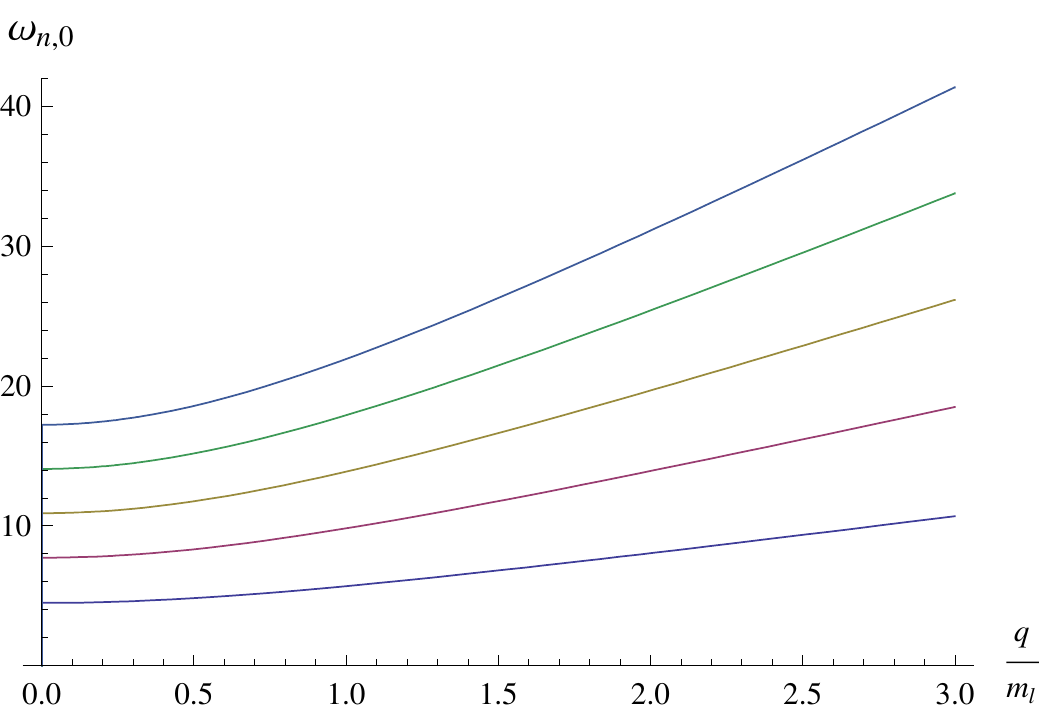}\;
    \epsfbox{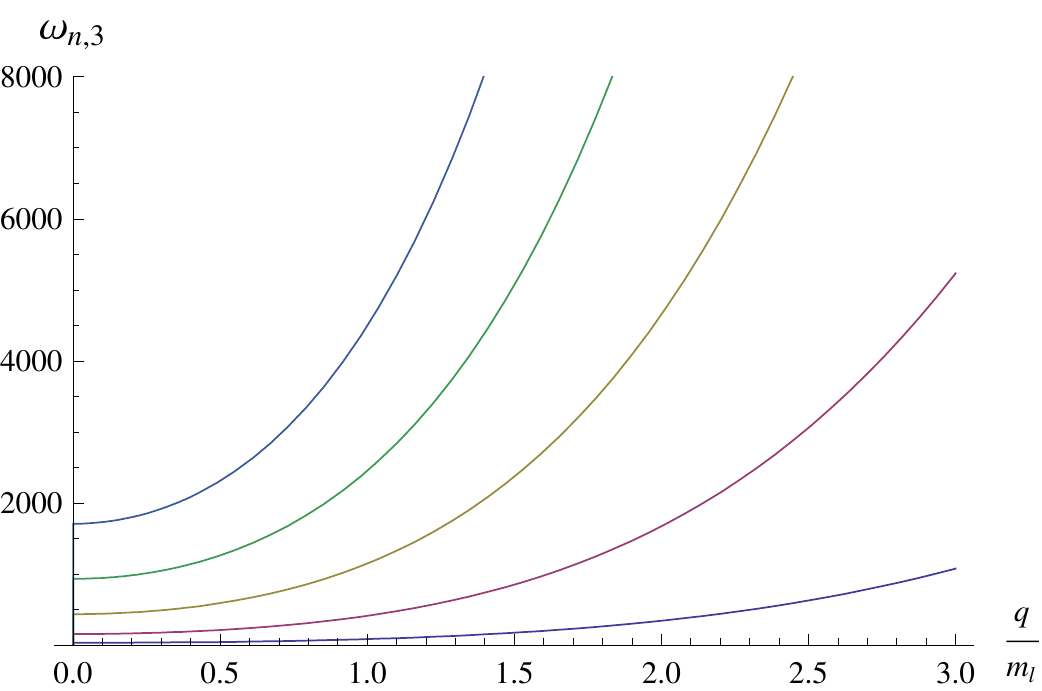}
    }
\caption[a] {Plots of the functions $\omega_{n,0}(q/m_l)$ and $\omega_{n,3}(q/m_l)$, respectively. The index $n$ grows from 1 to 5 from the bottom to the top curve in both figures.} }
}

For the $\delta\rho$ fluctuation spectrum given by Eq.~(\ref{rhomodes}), we have to resort to numerics. In the limit of large $y_h\gg y_l$, we may use our earlier numerical solution utilizing a power series expansion in $y_l/y_h$, in terms of which the 
spectrum
can be written in the form
\ba
E^\rho_n&=&m_h-m_l+ m_l\, \omega_{n,0}(q/m_l)\fr{2\pi}{\sqrt{\lambda}}+\fr{m_l^4}{m_h^3}\omega_{n,3}(q/m_l)\fr{2\pi}{\sqrt{\lambda}}+{\mathcal O}(m_l^5/m_h^4)\ .
\ea
This formula corresponds to the energy of a string with a single quantum of energy in the $n$th mode
of the $\rho$ fluctuations.
We plot the functions $\omega_{n,0}$ and $\omega_{n,3}$ in Fig.~\ref{fig:rhofluct}.
From there, we see that the energies of the fluctuations are always minimized
at $\rho_0=0$ or $q=0$, just as it was for the $x$ and $y_6$ fluctuations.
Another interesting aspect of these excitation energies is the absence of the two first leading corrections in $m_l/m_h$ in the heavy quark limit.


\section{Spinning strings}
\label{sec:spinning}

To supplement our discussion of the small fluctuations of strings around static quark-antiquark solutions, we now turn to consider the case where the string joining the heavy and the light brane is spinning. First, we consider strings spinning in the real space where they have a conserved angular momentum, and then look into strings spinning in
the internal $\theta$ direction where the corresponding angular momentum can be reinterpreted as a charge.
Our analysis is purely classical, but we expect valid, provided the angular momentum and
charge of the strings are large.

As we have discussed briefly already, there is an interesting wrinkle in the discussion of the
$\rho$--$\theta$ spinning strings.  A straight, motionless string stretching between the D7 branes
at a nonzero value of $\rho_0$ is a solution for all $\rho_0$.  That such a string is a solution is
surprising given the lack of translation invariance in $\rho$.  As we saw before in the analysis
of the fluctuations, if we excite one of these straight strings with $\rho_0 \neq 0$, it will experience
a force pulling it toward $\rho=0$.  In this section on spinning strings, we will find that
a string spinning in the $\rho$--$\theta$ plane is not free to sit at an arbitrary average value
of $\rho_0$ either.

\subsection{Strings spinning in real space}

We start by looking into the profile and energy spectrum of a string spinning in real space, more specifically in the  $x^1$--$x^2$ plane, assuming
that $x^3=\rho=y_6=0$. To begin with, we transform from Cartesian $(x^1, x^2)$ to polar coordinates $(r,\phi)$, and make the uniformly rotating ansatz of  Ref.~\cite{Kruczenski:2003be},
where $\phi = \Omega t$ is independent of
the worldsheet coordinate $\sigma$.  At the same time, we assume that $z(\sigma)$ and $r(\sigma)$ are
$t$ independent, which leads to an action of the form
\be
S = - \frac{L^2}{2 \pi \alpha'} \int dt \, d\sigma \, \frac{1}{z^2} \sqrt{(1-\Omega^2 r^2)((z')^2 + (r')^2)} \ ,
\ee
invariant under reparametrizations of the worldsheet coordinate $\sigma = f(\sigma')$.
For the most part, we will choose $\sigma = z$, though for the numerical studies we will shortly
present, we found it sometimes convenient to make other choices, such as $\sigma = r$.
This action leads to the following formulae for the energy and angular momentum of the string:
\ba
E &=& \frac{L^2 }{2 \pi \alpha'} \int d\sigma \, \frac{1}{z^2} \sqrt{\frac{(z')^2+(r')^2}{1-\Omega^2 r^2}} \ ,\label{srE} \\
J &=& \frac{L^2 \Omega }{2 \pi \alpha'} \int d\sigma \, \frac{r^2}{z^2} \sqrt{\frac{(z')^2+(r')^2}{1-\Omega^2 r^2}} \label{srJ} \ .
\ea

Choosing now $\sigma=r$, the equation of motion for $r(z)$ has the form
\be
\frac{r''}{1+(r')^2} - \frac{2}{z} r' + \frac{\Omega^2 r}{1-\Omega^2 r^2} = 0 \ , \label{reom}
\ee
which we now proceed to solve, demanding that Neumann boundary conditions be satisfied on the heavy and light branes at $z=z_h$ and $z=z_l$.  Neumann boundary conditions for $\phi$ are
satisfied trivially because $\phi'=0$, while for $r$ the boundary conditions read
\be
\left . r' \sqrt{\frac{1-\Omega^2 r^2}{1+(r')^2} } \right|_{z=z_h,z_l}= 0 \ .
\ee
Thus, we must either require that $r'=0$ at the boundary or that $\Omega^2 r^2=1$, which
physically is the condition that the endpoint of the string is moving at the local speed of light.
We will in general choose $r'=0$, but will nevertheless find certain ``critical'' solutions that satisfy
the light-like boundary conditions.

The linearized form of Eq.~(\ref{reom}) provides a good place to begin our study, as this form
\be
z^2 \left( \frac{r'}{z^2} \right)' + \Omega^2 r = 0 \ ,
\label{linearized}
\ee
of Eq.~(\ref{reom}), valid when $r'$ and $\Omega r \ll 1$,
is easy to solve.  Indeed, we already solved it; Eq.~(\ref{linearized}) is identical
to Eq.~(\ref{eomone}) in the case $\rho_0=0$.
Assuming then that the string takes the form
\ba
r &=& A\left( \omega_n z \cos(\omega_n(z-z_l)) - \sin(\omega_n(z-z_l)) \right) , \\
\phi &=& \omega_n t = \frac{\pi n}{z_l-z_h} t\
\ea
for small $A$, where we have adapted Eq.~(\ref{xfluctzero}), the energy and angular momentum are given by the approximate expressions
\ba
E &= & \frac{L^2}{2\pi\alpha'} \left( \frac{1}{z_h} - \frac{1}{z_l} - \frac{(\pi n)^4 A^2}{2 (z_h-z_l)^3}
+ {\mathcal O}(A^4) \right) \ , \\
J & = & \frac{L^2}{2 \pi \alpha} \left(\frac{(\pi n )^3 A^2}{2(z_h-z_l)^2} + {\mathcal O}(A^4) \right) \ .
\ea
Eliminating $A$ from here, we find that\footnote{The $n=1$ version of this formula (\ref{smallJ}) was first presented in Ref.~\cite{ParedesTalavera}.}
\be
E \approx m_h - m_l + n\pi \frac{m_l m_h}{m_h-m_l} \frac{2 \pi J}{\sqrt{\lambda}}, \label{smallJ}
\ee
which corresponds to the dashed straight lines in Fig.~\ref{fig:EJplot} (left), where we display the $E$ vs.~$J$ dependence of our spinning strings.
This linear scaling of $E$ with $J$ is characteristic of a particle in a Hooke's law potential, where the
constant of proportionality is given by the frequency of the oscillator.


\FIGURE[t]{
\label{fig:spinning1}
{ \centerline{\def\epsfsize#1#2{1.3#1}
    \epsfbox{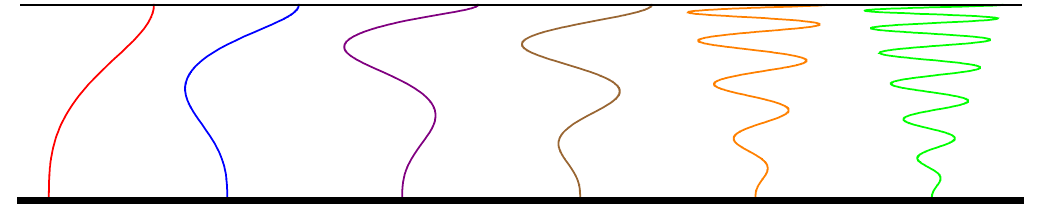}}
    \\
  \centerline{\def\epsfsize#1#2{1.3#1}
    \epsfbox{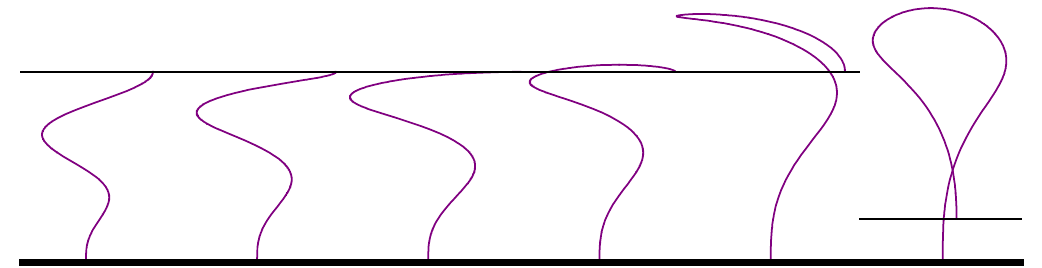}}

\caption{Top: A schematic plot showing the forms of the spinning string solutions $r(z)$ for various $n$. The lower (thick) horizontal line corresponds to the heavy brane sitting at $z_h=1/100$ and the upper (thin) line to the light brane at $z_l=1$, with the coordinate $z$ growing vertically. The six curves, from left to right, correspond to the cases of $n=1,2,3,4,9$ and 13, respectively.
Bottom: Another schematic plot showing the evolution of the $n=3$ branch as $\Omega$ is decreased
from 9.52 (left) to 0.5 (right). The critical solution $\Omega_{3c}=5.84$ is the third from the left. For the smallest value of $\Omega$, corresponding to large $J$ and $E$, we have rescaled the solution in the $z$ direction by a factor of 4.4 in order to make it fit in the figure. In the $\Omega \rightarrow 0$ limit, the solution becomes symmetric in the $r$ direction about the center of mass.}
}}


As the $E$ and $J$ of the string get larger, $r$ will get larger as well, and eventually our
linearized approximation breaks down.  To make further progress, we resort to numerics
to calculate the profile $(r, z)$ of the spinning strings.
For simplicity, we rescale our variables so that $z_l=1$, and have $z_h$ take the values $1/10$ and $1/100$, corresponding roughly to the heavy-to-light quark mass ratios one finds in QCD for charm and bottom quarks. We find that for each $n$, there is a continuous family of rotating string solutions for all $\Omega$ such that $0 < \Omega < \omega_n$. The index $n$ parametrizes the number of turning points in the solutions: For the branch $n$, the string profile $(r,z)$ has always $n-1$ (local) extremal values in $r$. Examples of the profile $(r, z)$ for various $n$ are exhibited in Fig.~\ref{fig:spinning1} (top).

Once the results for $(r, z)$ are obtained in a numerical form, we insert them into the integrals of Eqs.~(\ref{srE}) and (\ref{srJ}), thus obtaining the energies of the spinning strings in terms of their angular momenta. The resulting curves $f(x)$, parametrizing the energies through
\ba
E&=&m_h-m_l+m_l\,  f(2\pi J/ \sqrt{\lambda}),
\ea
are shown for $n=1,2,3,4$ and $z_h=1/100$ in Fig.~\ref{fig:EJplot} (left) and in more detail for the $n=1$ branch in Fig.~\ref{fig:EJnzeroplot}.
Intriguingly, reducing $\Omega$ increases
both $E$ and $J$.  A similar behavior was observed for the heavy-heavy mesons in
Ref.~\cite{Kruczenski:2003be}, and is explained by the fact that
the decrease in $\Omega$ is made up for by the growing size of the string.
The evolution of the profile of the $n=3$ branch string as a function of $\Omega$ 
is shown in Fig.~\ref{fig:spinning1} (bottom).


\FIGURE[t]{
\label{fig:EJplot}
\centerline{
 \epsfig{figure=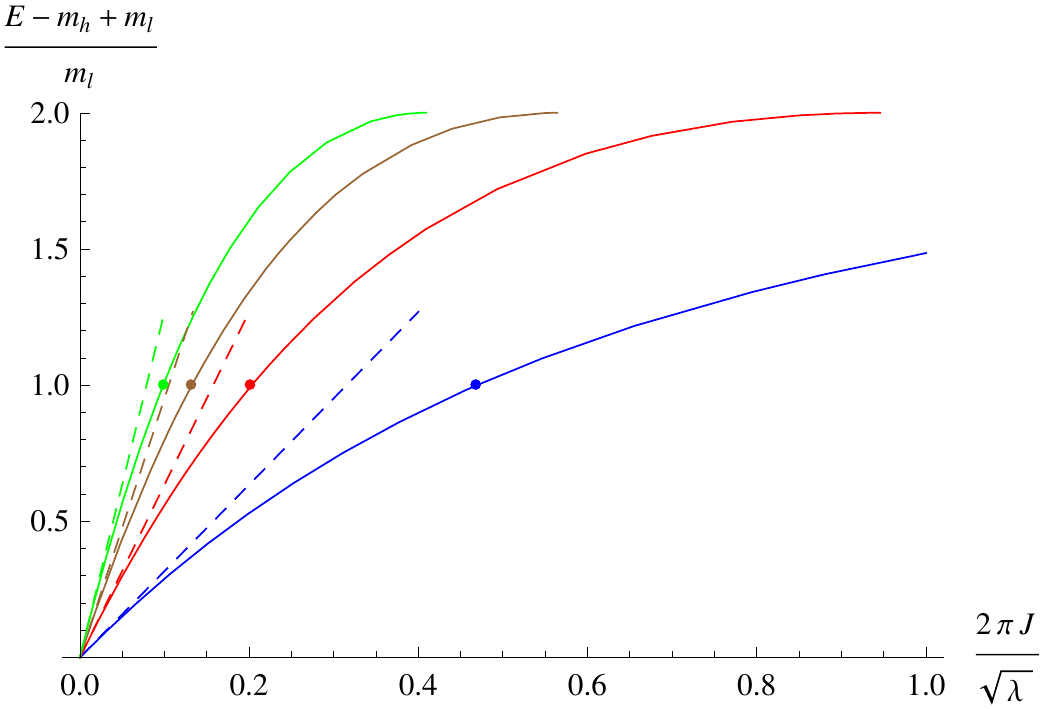, width=3in}
 \epsfig{figure=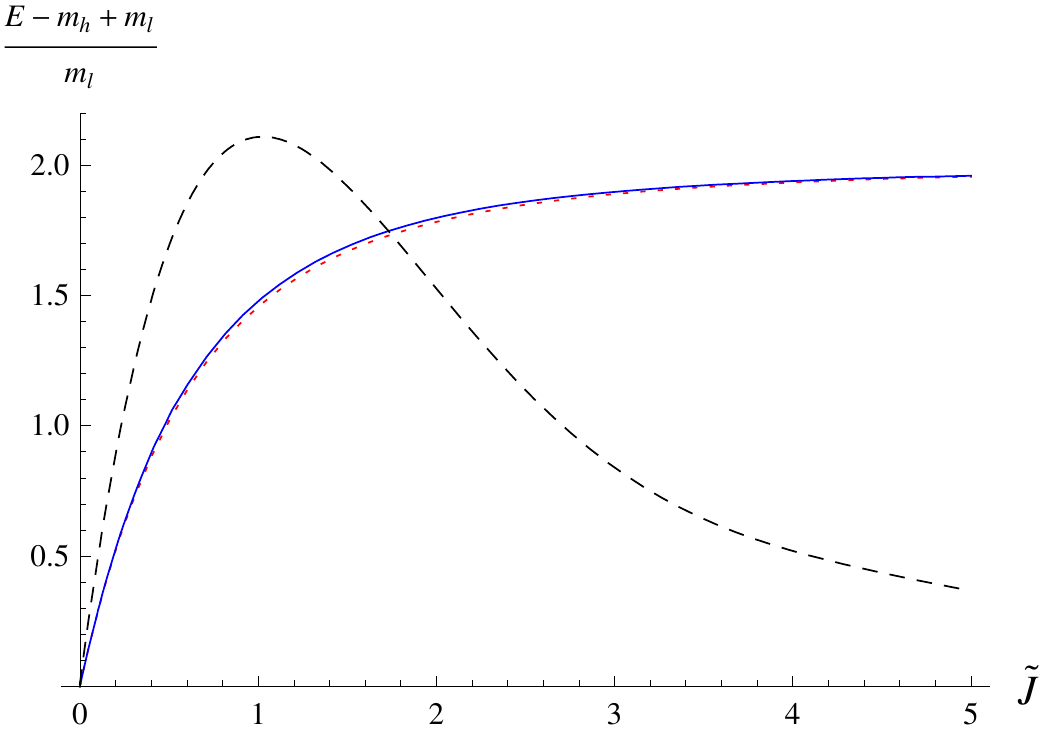, width=3in}
}
\caption{Left: The dependence of $E$ versus $J$ for the spinning heavy-light mesons. We display the curves for $m_h=100m_l$ and $n=1,2,3,4$ from right to left, with the adjacent dashed straight lines corresponding to the respective analytic small-$J$ approximations of Eq.~(\ref{smallJ}) and the dots on the curves denoting the critical solutions at $\Omega=\Omega_{nc}$. Right: The $E(\tilde J)$ curves for both the $m_h=100m_l$ (solid blue curve) and $m_h=10m_l$ (dotted red) cases for the $n=1$ branch, together with their difference multiplied by a factor of 100 (dashed black).}
}


The dependence of the $E(J)$ curves on $m_h$ is relatively mild and easily modeled.
The Eq.~(\ref{smallJ}) suggests a rescaling  of the variable $J$ by $1/(1-m_l/m_h)$, defining
\ba
\tilde J &=& \frac{m_h}{m_h-m_l}\fr{2\pi J}{\sqrt{\lambda}}.
\ea
With this small correction, we see from Fig.~\ref{fig:EJplot} (right) that the curves corresponding to $z_h=1/10$ and 1/100 practically overlap.

As $\Omega$ is decreased, there is a critical $\Omega_{nc}$ for each family of solutions where
the light quark endpoint of the string is moving at the local speed of light, $\Omega_{nc}\, r(z_l) = 1$.
For the {\it short strings}
with $\Omega > \Omega_{nc}$, the string is contained entirely between the two D7 branes, while
for the {\it long strings} with $\Omega < \Omega_{nc}$, there is a loop of string in the region $z > z_l$.
Like the $\omega_n$,
the critical $\Omega_{nc}$ depend to some extent on the choice of the heavy and light quark masses.
For the first few $n$, we find that
\ba
z_h&=&1/10:\; \Omega_{1c} = 1.54,\; \Omega_{2c}=3.98,\;\Omega_{3c}=6.22,\;\Omega_{4c}=8.41,\;\nn
z_h&=&1/100:\; \Omega_{1c} = 1.38,\; \Omega_{2c}=3.72,\;\Omega_{3c}=5.84,\;\Omega_{4c}=7.91. \nonumber
\ea
We furthermore observe that for $n=1$, the critical energies and angular momenta obey the results
\ba
E_c&=&m_h-\fr{m_l^2}{2m_h}+{\mathcal O}\(\fr{m_l^3}{m_h^2}\) ,\;\\
\tilde{J}_c&=&0.473-0.262\,\fr{m_l}{m_h}+{\mathcal O}\(\fr{m_l^2}{m_h^2}\),
\ea
and that for $n>1$, the forms of the equations stay intact, while the numbers in the latter relation somewhat change.
Especially the former of these results deserves some attention; we have verified this relation to more than 1 part in 10000, but have so far no explanation for why the limiting energy should obtain such a simple form.


\FIGURE[t]{
\label{fig:EJnzeroplot}
\centerline{
\epsfig{figure=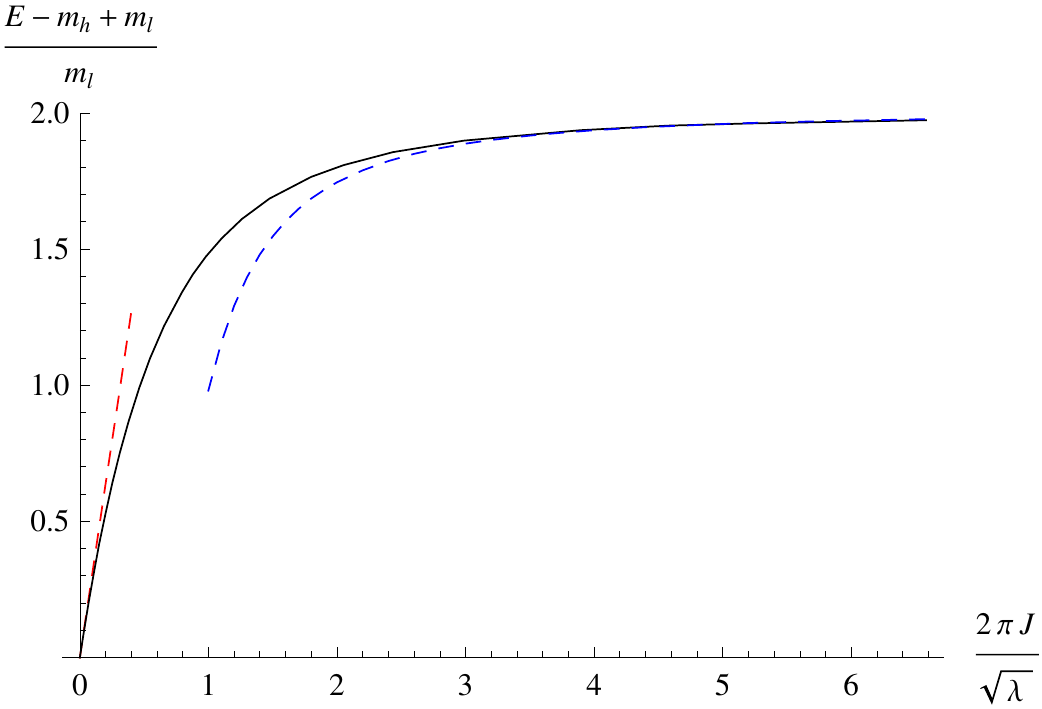, width=3.5in}
}
\caption{
We plot $E$ versus $J$ for the $n=1$ branch of the spinning heavy-light mesons.
The solid curve is the numerical result for
the case $m_h = 100 m_l$, while the red and blue dashed curves are the analytic small and large-$J$
approximations of Eqs.~(\ref{smallJ}) and (\ref{largeJ}), respectively.
}
}


As $\Omega$ is decreased below $\Omega_{nc}$, the strings quickly begin to get very large compared
to the separation between the D7 branes, and in the $\Omega\to 0$ limit, their size in fact diverges 
both in the $r$ and $z$ directions. Indeed, in this limit the spinning string solutions can be seen to approach
those of the heavy-heavy mesons considered in Ref.~\cite{Kruczenski:2003be}, where both
ends of the string end on the same D7 brane. The limit $\Omega \to 0$ of the $n=1$ branch is special because
the velocity of any point on the $n=1$ string approaches zero as $\Omega$ decreases, while for
the $n>1$ branches, there always exists a finite set of points $\sigma_i$ along the string
where, due to the large size of the string, $r(\sigma_i) \Omega \to 1$ as $\Omega \to 0$.
As noticed originally by Refs.~\cite{Kruczenski:2003be, ParedesTalavera}, the small
size of $\Omega r$ allows for an analytic treatment of the $E$ and $J$ 
of the $n=1$ branch in the $\Omega \to 0$
limit.

In the $\Omega \to 0$ limit, the strings correspond to marginally bound heavy-light
mesons with an energy $E \approx m_h + m_l$.  By marginally bound, we mean that
the binding energy becomes very small.
For the $\Omega \to 0$ limit of the $n=1$ branch, the string profile must be well approximated
by the static configuration that determines the potential between two infinitely massive quarks.
As shown in Ref.~\cite{ParedesTalavera}, in this limit the energy of the string obeys the relation
\ba
E &=& m_h + m_l - \kappa \, \frac{m_l m_h}{m_h+ m_l} \frac{\lambda }{ J^2} \ ,
\label{largeJ}
\ea
where
\[
\kappa = 2  \left( \frac{\Gamma(3/4)}{\Gamma(1/4)} \right)^4\approx  0.0261\ ,
\]
consistent with a Coulombic attraction between the quarks.
We see from Fig.~\ref{fig:EJnzeroplot} that Eq.~(\ref{largeJ}) is quite a good approximation to
the $E(J)$ curve already at moderately large $J$.
In contrast, the $\Omega \to 0$ limit of the $n>1$ branches all terminate at finite values of $J$.
Numerically, for the case of $m_h = 100 m_l$, these terminal values of $2 \pi J / \sqrt{\lambda}$ are
$0.946$, $0.546$, and $0.409$ for the $n=2$, 3 and 4 branches, respectively.

We believe that the long strings are much less stable than the short strings. For one, they
intersect the D7 brane and thus can break in two. For another, they are much bigger in size than the short strings,
and thus it is likely that they are subject to instabilities, which do not respect the
uniformly rotating $\phi = \Omega t$ ansatz.

\subsection{String profile in $\rho$ and $\theta$}

Next, we look at the profile of a string spinning inside the ${\mathbb R}^6$, in the $\rho$--$\theta$ directions. Let $Q$
be the corresponding angular momentum.
Although $Q$ is an angular momentum from the ten dimensional point of view, in the
four dimensional field theory it is a charge, namely the R-charge of the
R-symmetry of our supersymmetric field theory. From the point of view of QCD, $Q$ could be viewed as
a model of the electromagnetic charge of the meson.

To begin with, we assume that $x=r=y_6=0$, and in analogy with our discussion of strings spinning in real space, make an ansatz where
$\rho(y)$ is time independent and $\theta = \Omega t$ is $y$ independent.
The Neumann boundary conditions for $\theta$ are then again trivially satisfied because $\theta'=0$.
With these simplifications, the action for the string reduces to
\ba
S_{NG} &=& -\frac{L^2}{2\pi\alpha'} \int dt \, dy \, \sqrt{\(1-\rho^2\Omega^2 /u^4 \)\(1+(\rho')^2\)} \ ,
\ea
leading to the equation of motion for $\rho(y)$,
\ba
\frac{u^2 \rho''}{1+(\rho')^2}+\Omega^2 \rho \frac{u^2-2\rho^2+2y\rho\rho'} {u^4-\Omega^2\rho^2}&=&0 \ . \label{eomrho2}
\ea

The energy $E$ and internal angular momentum $Q$ of the spinning strings are given by
\ba
\label{energy}
E &=& \frac{L^2}{2 \pi \alpha'} \int dy \sqrt{\frac{1+ (\rho')^2}{1- \rho^2 \Omega^2/u^4}} \ ,  \\
Q &=& \frac{L^2}{2\pi\alpha'} \int dy \frac{\rho^2 \Omega}{u^4} \sqrt{\frac{1+ (\rho')^2}{1- \rho^2 \Omega^2/u^4}} \ .
\label{Qch}
\ea
The Neumann boundary conditions for $\rho$ on the other hand reduce to the requirement
\be
\left. \rho' \sqrt{\frac{1-\rho^2 \Omega^2/u^4}{1+(\rho')^2}} \right|_{y=y_h,y_l}=0 \ ,
\ee
from where we see that we must again either require that $\rho'=0$ at the boundary or that the ends of the string move
at the local speed of light. Similar to the strings spinning in real space, we
generically enforce $\rho'=0$, but in addition find certain special solutions that satisfy the light-like
boundary conditions.
Note that a motionless string with $\rho = \rho_0$ and $\Omega = 0$ is a solution to the equations
of motion for all $\rho_0$.  Once $\Omega \neq 0$, however, the story becomes much more interesting.

For non-zero $\Omega$, the equation of motion for $\rho$, Eq.~(\ref{eomrho2}), seems difficult to solve analytically at least in full generality, and we will
therefore resort to numerics, setting again $y_l=1$ and varying the location of the heavy brane $y_h$.
The story we encounter is strongly reminiscent of the strings spinning in real space.
We again find multiple branches of solutions indexed by an integer $n$, $n\geq 1$, with the corresponding string profiles $\rho_n(y)$ containing exactly $n-1$ extrema in $\rho$.

The low energy behavior of our strings can again be understood analytically through the fluctuation
analysis of the previous Section. In this $E\rightarrow 0$ limit, we may take the string profiles to be complex combinations of
$\rho$ fluctuations with infinitesimal amplitude. The complex combination produces a string spinning in the $\rho$--$\theta$ plane
with angular velocity $\Omega = \omega_n$, corresponding to the solutions to Eq.~(\ref{rhozero}).
Consistent with the results from Section 4.3, we see that for $y_h=100$, the values of the first few $\omega_n$'s are 4.493, 7.725, 10.904.


\FIGURE[t]{
\label{intspaceprofiles}
{ \centerline{\def\epsfsize#1#2{0.8#1}
    \epsfbox{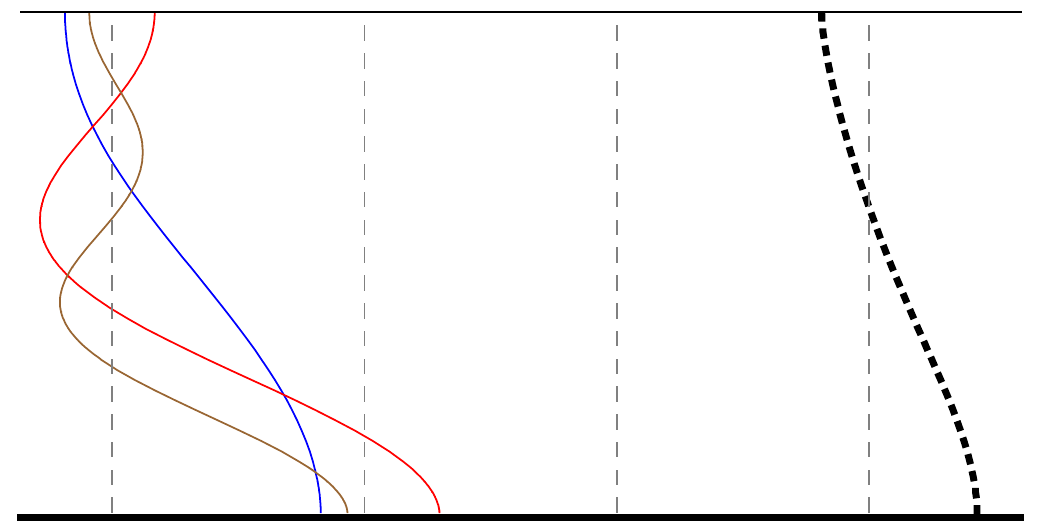}}
    \centerline{\def\epsfsize#1#2{1.3#1}
    \epsfbox{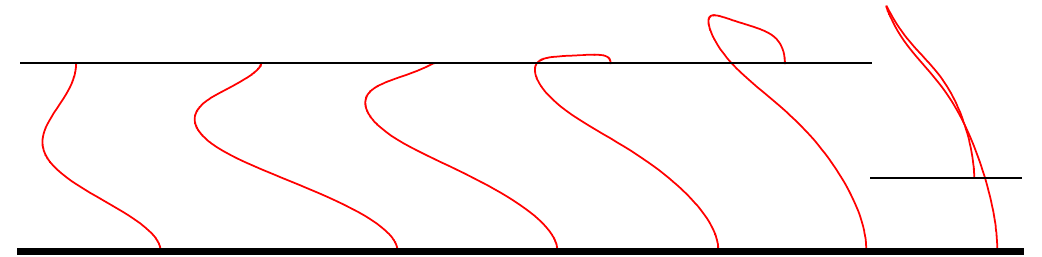}}
\caption{Top: Profiles of the spinning strings $\rho(z)$ stretching between the two branes at $z_l=1$ and $z_h=1/100$, with $z\equiv 1/y$ and the notation as in
Fig.~\ref{fig:spinning1}. The black dotted line corresponds to the $n=0$ branch, while the blue, red and brown solid curves correspond to the $n=1,2,3$ cases, respectively. The gray dashed lines, from left to right, mark the points $\rho=0,1/2,1,3/2$.
Bottom: The evolution of the $n=2$ branch of solutions as $\Omega$ is decreased from $7.725$ to 1. The critical solution is again the third from the left, while the smallest $\Omega$ solution has been rescaled in the $z$ direction by a factor 2.55. In the $\Omega\to 0$ limit, the part of the solution extending beyond the light brane doubles back on itself.}
}}


For a given $n>0$, we find a continuous family of solutions in the range $0< \Omega < \omega_n$.
Decreasing $\Omega$ corresponds to increasing $E$ and $J$, the increase in the size
of the string more than making up for the loss of angular velocity.
There are again critical angular frequencies $\Omega_{nc}$ which separate the {\it long strings} with $\Omega < \Omega_{nc}$
from the {\it short strings} with $\Omega > \Omega_{nc}$, the former extending to the region $y<y_l$.
For the critical solution, the endpoint of the string sitting
on the light brane is moving at the local speed of light. 
For $y_h=100$, the critical angular velocities for the first three branches are found to equal $\Omega_{1c}=3.260$, $\Omega_{2c}=5.152$ and $\Omega_{3c}=7.108.$

In addition to the branches with $n\geq 1$, we find an additional branch of solutions, $n=0$, which has no analog for the strings spinning
in real space.
This branch of the spinning strings emerges from the lifting
of the zero fluctuation mode corresponding to translations in the $\rho$ direction, and as we will show  shortly, it is
possible to understand its low-energy properties in a semi-analytic fashion.
Earlier in our fluctuation analysis,
we saw that while the ground state string sitting at $\rho \neq 0$ with $\Omega=0$ did not
experience a potential, excited strings felt a force pulling them toward $\rho=0$.
Here, we instead find that strings with even an arbitrarily small $\Omega$ are not free to move in
the $\rho$ direction, but must
sit at a constant $\rho=\rho_0$ in the limit where $\Omega$ tends to zero.

Inspecting the $n=0$ branch numerically for $y_h=100$, we observe that $\Omega$ can be arbitrarily close to zero, with the $\Omega \to 0$
limit corresponding to small angular momenta and energies, in contrast to the branches with $n \geq 1$. In this limit, the string profile becomes
a constant, equaling $\rho(y)\equiv \rho_0\approx 1.825$. This time there is no maximal angular velocity at which the solution breaks down,
but we rather find that the curve that this branch of solutions draws
on the ($\Omega,\,\rho(y_l)$) plane is not a single valued function of $\Omega$. For the $y_h=100$ case we are considering, it starts from the point (0,\,1.825), follows monotonically
to the point (2.082,\,1.361) and finally turns back to end at (2.069,\,1.300), where the light end of the string is spinning at the local speed of light. We exhibit the forms of the string profiles for $n=0,1,2,3$ in Fig.~\ref{intspaceprofiles}.


\FIGURE[t]{
\label{fig:EQplot}
\centerline{
\epsfig{figure=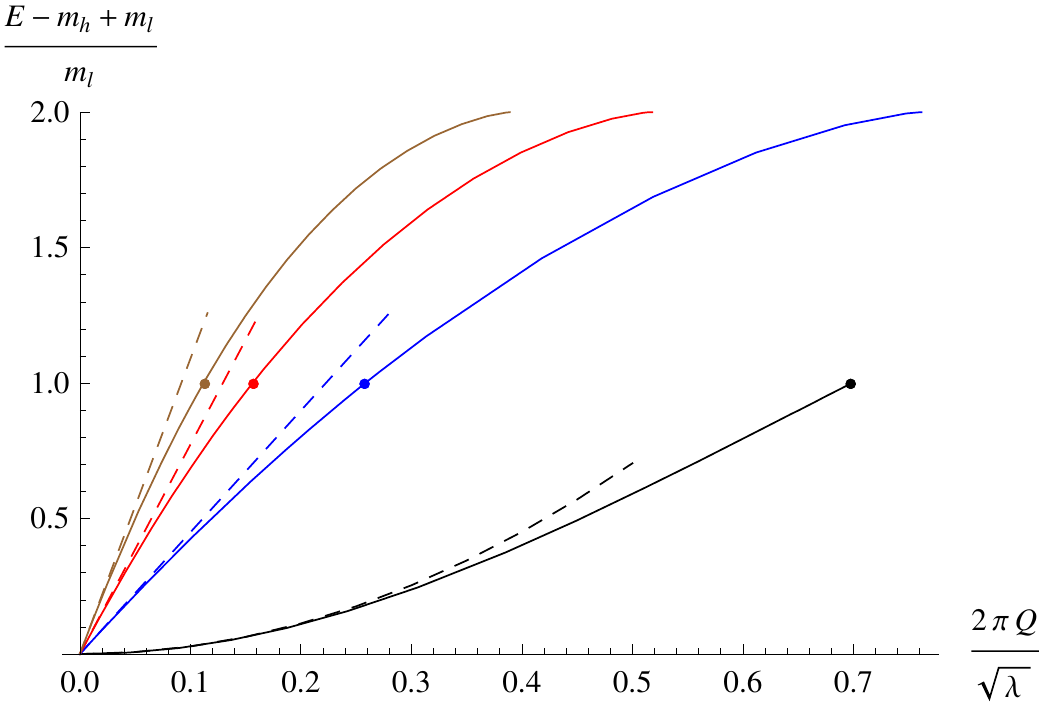, width=3in}\;\;
\epsfig{figure=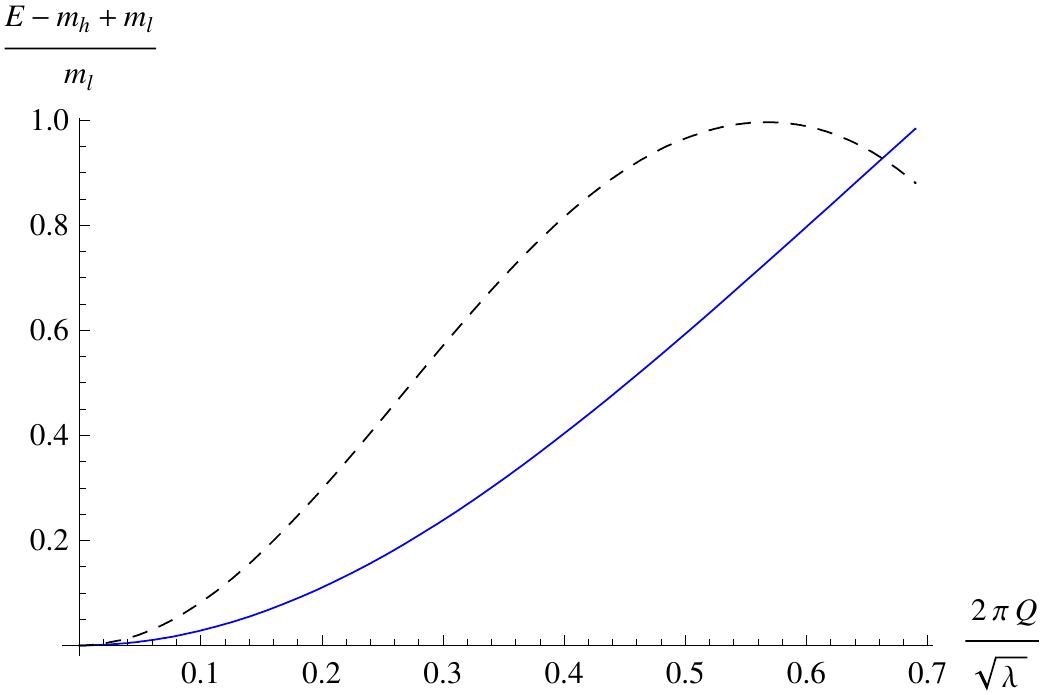, width=3in}
}
\caption{Left: $E$ versus $Q$ for the spinning heavy-light mesons.  From right to left, the solid curves correspond to the $n=0,1,2,3$ branches, and the corresponding dashed curves to the analytic small $Q$ approximations of Eqs.~(\ref{appr1}) and (\ref{analyticEQ}). The value of $y_h$ is set to 100, and the dots on the curves again denote the critical solutions. Right: The effect of changing the heavy brane location from $y_h=100$ (solid blue curve) to $y_h=10$ (dotted red) in the $n=0$ case. The difference of the two curves is also shown as the dashed black line, magnified by a factor of 500.}
}


In Fig.~\ref{fig:EQplot}, we plot the $E$ vs.~$Q$ dependence of the different branches of spinning string solutions we have encountered.
Let us first focus on the $n \geq 1$ branches, and specifically on their small $Q$ limits.
Similar to the analysis of the strings spinning in real space, we can consider the
approximate solution, valid for small $A$,
\ba
\delta \rho &=& A \frac{1}{z} \left( \omega_n z_l \cos (\omega_n (z-z_l)) + \sin(\omega_n (z-z_l)) \right) \ , \\
\theta &=& \omega_n t \ ,
\ea
with $z\equiv 1/y$ and the $\omega_n$'s given by our $\rho$ fluctuation spectrum. This solution leads to the approximate small $Q$ relation
\be
E \approx m_h - m_l + m_l \, \omega_n z_l \, \frac{2 \pi Q}{\sqrt{\lambda}}  \ , \label{appr1}
\ee
which is shown as the dashed straight lines in Fig.~\ref{fig:EQplot} (left).

Decreasing $\Omega$ towards the critical angular velocities $\Omega_{nc}$, $n\geq 1$, we observe that the charge $Q$ approaches a critical value $Q_{nc}$, varying according to $n$, while the energy $E$ approaches a universal constant $E_c\approx m_h$, independent of the branch in question. Both values, as well as the forms of the $E(Q)$ curves, are highly independent of the location of the heavy brane at sufficiently large values of $y_h$, and
for $y_h\geq 10$, the first few values of $Q_{nc}$ are $Q_{1c} = 0.258 \sqrt{\lambda} / 2\pi$,
$Q_{2c} = 0.156 \sqrt{\lambda} / 2\pi$, and $Q_{3c} = 0.112 \sqrt{\lambda} / 2\pi$.
This $m_h$ independence can be understood by inspecting the form of the canonical momentum densities appearing in Eqs.~(\ref{energy})--(\ref{Qch}).  The charge density $\pi_\theta^0$
behaves at large $y$ as $1/y^4$.
The energy density scales at leading order as $\sqrt{\lambda}$, giving rise to the ground state mass
$m_h-m_l$ of the heavy-light meson, but the first correction also behaves as $1/y^4$.   These $1/y^4$
terms mean that the excitation energy as a function of the charge of the spinning string is highly insensitive to the form of the string profile at $y\gtrsim 10y_l$.

If we proceed to even smaller frequencies, $0 < \Omega < \Omega_{nc}$, we notice that these $n>0$ branches persist all the way down to zero.  In the limit $\Omega \to 0$, the strings become marginally bound, like their real-space spinning counterparts,
with an energy $E \approx m_h + m_l$.  In contrast, the charges $Q$ for the terminal solutions
are not universal. For the case $m_h = 100 m_l$, we find that
the terminal values of $2\pi Q/ \sqrt{\lambda}$ are 0.762, 0.518, and 0.390 for the $n=1$, 2, and
3 branches respectively.  Like their real-space spinning counterparts,
we suspect that these long strings are not stable for the exact same reasons.

Switching then to following the $n=0$ branch on the $(Q,E)$ plane, we observe that for a given value of the charge, these strings are always energetically favored in comparison with their $n\geq 1$ counterparts. In the limit $y_h \to \infty$, we find that the energy and charge of the critical solution on the $n=0$
branch very quickly approach
\ba
E_{0c} &=& m_h - 6(1)\,\frac{m_l^4}{m_h^3}  + {\mathcal O} \left( \frac{m_l^5}{m_h^4} \right),\\
\frac{2 \pi Q_{0c}}{\sqrt{\lambda}} &=& 0.69868(1)  -4.0(5)\,\frac{m_l^3}{m_h^3}+ {\mathcal O} \left( \frac{m_l^4}{m_h^4} \right)  ,
\ea
where the coefficients of the first terms have been found by fitting a variety of trial functions to our numerical data
and the errors have been estimated in a very conservative manner.
The vanishing of the first few corrections in $1/m_h$
is similar to the suppression of $1/m_h$ corrections
in the $\rho$ fluctuation analysis of Section \ref{sec:rhofluct}. The $n=0$ branch does not appear to admit long string solutions.

\subsubsection*{Small $\Omega$ limit of the $n=0$ branch}

To conclude our inspection of the string spinning in the $\theta$ direction, we will now take a closer look at the limit of infinitesimally small $\Omega$ in order to gain more understanding of the behavior of the $n=0$ solutions there. We note that this limit corresponds to approximating $y_l \gg \Omega$,
and therefore implies that we may use the relation $u^4-\Omega^2 \rho^2\approx u^4$ in the equation of motion for $\rho$. On the other hand, the observed fact that $\rho$ is nearly a constant in this case implies that
\ba
\(u^2-2\rho^2+2y_5\rho\rho'\)\(1+(\rho')^2\)\rho &\approx& \(u^2-2\rho^2\)\rho,
\ea
finally giving as the equation to solve
\ba
u^6 \rho''+\Omega^2\(u^2-2\rho^2\)\rho&=&0.
\ea

In the last form, we note that we may write
\ba
\rho(y)&=&\rho_0+\delta {\rho}(y),
\ea
where $\rho_0$ is a constant and $\delta {\rho}(y)$ satisfies the Neumann boundary conditions at $y=y_l$ and $y=y_h$.  We define $\rho_0$ by the constraint that $\delta{\rho}\rightarrow 0$, as
$y\rightarrow y_h$. Using this parametrization and the fact that $y_l\gg \Omega$, we see that $\delta{\rho}$ satisfies the equation of motion
\ba
\delta {\rho}'' &=&-\Omega^2\fr{\(y^2-\rho_0^2\)}{(y^2+\rho_0^2)^3}\rho_0 \ .
\ea
If we enforce the boundary condition at $y=y_h$, this differential equation can then be integrated to yield
\ba
\label{rhoapproxsol}
\delta {\rho}(y)/\Omega^2&=&\fr{(y-y_h)\(\rho_0^4(y-2y_h)-\rho_0^2 y y_h(3y-y_h)-y^2 y_h^3\)}{4\rho_0(y^2+\rho_0^2)(y_h^2+\rho_0^2)^2}\nn
&+&\fr{y}{4\rho_0^2}\(\arctan\bigg[\fr{y}{\rho_0}\bigg]-\arctan\bigg[\fr{y_h}{\rho_0}\bigg]\),
\ea
from where --- demanding that the derivative of this expression vanish also at $y=y_l$ --- we finally obtain as the equation for $\rho_0$
\ba
\label{rhozerocond}
\fr{(y_l^2+3\rho_0^2)y_l\rho_0}{(y_l^2+\rho_0^2)^2}-\fr{(y_h^2+3\rho_0^2)y_h\rho_0}{(y_h^2+\rho_0^2)^2}&=& \arctan\bigg[\fr{y_h}{\rho_0}\bigg]-\arctan\bigg[\fr{y_l}{\rho_0}\bigg].
\ea
Solving this equation numerically produces two solutions, $\rho_0=0$ and $\rho_0 = F(y_h/y_l)\times y_l$, of which we can throw out the former, as it is not consistent with our assumption of a small $\delta \rho$ and furthermore leads to a vanishing angular momentum. The latter result, on the other hand, is a slowly varying function of $y_h/y_l$ for large values of this ratio, approaching in the $y_h/y_l\rightarrow \infty$ limit the result $\rho_0 \approx 1.82526 \, y_l$.  In contrast,
for $y_h \approx y_l$, $F(y_h/y_l) \approx 1$.


\subsubsection*{Properties of the small-$\Omega$ solution}

To get some feeling for the physical properties of the above solutions obtained for small $\Omega\ll y_l$, we will next compute their energy $E$ and internal angular momentum $Q$ using Eq.~(\ref{energy}),
where the canonical momentum and internal angular momentum densities read approximately
\be
\pi_t^0
\approx
- \frac{L^2}{2 \pi \alpha'} \left(1 + \frac{\rho_0^2 \Omega^2}{2 u_0^4} \right) \qquad \mbox{and} \qquad
\pi_{\theta}^0
\approx \fr{L^2}{2\pi \alpha'}\fr{\rho_0^2\Omega}{u_0^4},
\ee
with $u_0^2\equiv y^2+\rho_0^2$. Here, we have neglected higher order corrections in $\Omega$ and
used the approximate solution (\ref{rhoapproxsol}).
Performing the integrals, we obtain
\be
E \approx \fr{L^2}{2\pi \alpha'}\big(y_h-y_l+\fr{\Omega^2}{2 y_l} \Upsilon \big) \qquad \mbox{and} \qquad
Q \approx \fr{L^2}{2\pi \alpha'}\fr{\Omega}{y_l}\Upsilon, \label{Jomega}
\ee
in which we have defined the dimensionless constant
\ba
\label{upsilon1}
\Upsilon &\equiv& \rho_0^2 y_l \int_{y_l}^{y_h} dy\,\fr{1}{(y^2+\rho_0^2)^2}
= \rho_0^2 y_l \left(\frac{y_l}{(\rho_0^2+y_l^2)^2}   - \frac{y_h}{(\rho_0^2+y_h^2)^2} \right) \ .
\ea
In deriving Eq.~(\ref{upsilon1}),
we have made use of Eq.~(\ref{rhozerocond}).  Note that we have
\be
\lim_{y_h \to \infty} \Upsilon \approx 0.17757 \qquad \mbox{while} \qquad \lim_{y_h \to y_l} \Upsilon = \frac{y_h-y_l}{4 y_l} \ .
\ee

We may now easily solve $\Omega$ in terms of $Q$ from Eq.~(\ref{Jomega}) above, which allows us to write $E$ in terms of $Q$
\ba
E&\approx&m_h-m_l+\fr{m_l}{2\Upsilon} \(\fr{2\pi Q}{\sqrt{\lambda}}\)^2.
\label{analyticEQ}
\ea
Thus we find again that the excitation spectrum does not depend on $m_h$ at leading order
in the heavy
quark mass limit.  As we can see from Fig.~\ref{fig:EQplot}, this analytic approximation is quite
good even for moderately large values of $Q$.


\section{Summary and Discussion}

Although different in many respects,
the heavy-light mesons we have studied have a spectrum which shares certain properties
of real-world heavy-light mesons.  For example, consider the case where there are two heavy
quarks $h$ and $h'$ and two light quarks $l$ and $l'$.  We find for the ground state
heavy-light mesons that
\be
M_{hl} - M_{hl'} = m_{l'} - m_l = M_{h'l} - M_{h'l'} \ .
\ee
This kind of relation is similar to the real world relation (see for example Ref.~\cite{Neubert:1993mb})
for mesons containing a charm or bottom quark,
\be
m_{B_s} - m_B \approx m_{D_s} - m_{D} \approx 100 \mbox{ MeV} \ .
\ee

Of course, the sign of the above difference is wrong:  While for us, given that $m_{l} > m_{l'}$,
we would find a negative difference, in the real world the difference is positive.
This sign difference is, however, of little significance in this ${\mathcal N}=2$ SYM theory.
In Section \ref{sec:SUSY}, we noted that we could let
the lighter D7 brane end along $w^3 = c'$ where $c' \in {\mathbb C}$ and $|c'| =  1/z_l$.
This case still
preserves ${\mathcal N}=2$ supersymmetry and allows us to tune the mass of the ground state
heavy-light meson to be anything between $m_h - m_l$ and
$m_h + m_l$.  We did not study the excitation spectra of these more general
heavy-light mesons in this paper,
but it would be an interesting project for the future.

What we calculated was a portion of the heavy-light meson spectrum for hypermultiplets
with masses with the same phase.  In the dual language, both of our D7 branes
sit at $y_6=0$ (or equivalently $\mbox{Im} \, w^3 = 0$) and different values of $y_5$.
One generic feature of this spectrum is the $m_h$ independence of the excitation
energies in the heavy quark mass limit.
For example, for low lying fluctuations in the $x$ and $y_6$ directions we found the
energy spectrum
\be
E_n = m_h - m_l + m_l \frac{2 \pi^2 n}{\sqrt{\lambda}} + {\mathcal O}\left(\frac{m_l^2}{m_h} \right) \ .
\ee
For the $\rho$ fluctuations, we were not able to determine a spectrum analytically,
but were nevertheless able to determine this $m_h$ independence numerically.
The $x$ fluctuations should correspond to vector like mesons, while the $y_6$
and $\rho$ fluctuations should correspond to scalar like mesons.

We also studied spinning strings.  For the strings spinning in real space, we found
several branches, characterized by a radial excitation number $n$.  For small angular
momentum $J$, we were able to determine the analytic formula
\be \label{summaryEJ}
E = m_h - m_l + m_l \frac{2 \pi^2 n J}{\sqrt{\lambda}} + {\mathcal O}\left( \frac{m_l^2}{m_h} \right) \ ,
\ee
which displays this $m_h$ independence.
Finally we studied strings spinning in an internal space, which corresponds to mesons
with R-charge $Q$ from the field theory perspective.  For small $Q$, we found the
analytic formulae of Eqs.~(\ref{appr1}) and (\ref{analyticEQ})
which again displays $m_h$ independence.

Continuing the comparison with QCD,
we can consider the mass difference between an excited and a ground state heavy-light meson in QCD.
From the review \cite{Neubert:1993mb}, we learn that a typical QCD prediction
of this heavy quark limit is that the difference in energy between
excited and ground state heavy-light mesons should obey the relations
\be
m_{B_2^*} - m_B \approx m_{D_2^*}-m_D \approx 593 \mbox{ MeV} \; , \; \; \;
m_{B_1} - m_B \approx m_{D_1} - m_D \approx 557 \mbox{ MeV} \ .
\ee
Unfortunately, there is no good data yet for $m_{B_2^*}$ and $m_{B_1}$.
These differences are consistent with our result that the energy excitations scale with
$m_l$, although in real world QCD, we expect to have $m_l$ replaced with $\Lambda_{\rm{QCD}}$.

The electromagnetic mass splittings of heavy-light mesons in QCD are typically tiny
\cite{PDG}.  For
example, $m_{D^\pm} - m_{D^0} \approx 5$ MeV while
$m_{B^0} - m_{B^\pm} \approx 0.4$ MeV.  It is suggestive that in the large $\lambda$ limit,
our approximate formula (\ref{analyticEQ}) for the $Q$ dependence of the masses
is suppressed by an additional power of $Q/\sqrt{\lambda}$ compared with
the linear scaling of Eq.~(\ref{summaryEJ}) on $J/\sqrt{\lambda}$.  However,
we have no good understanding of the relative sizes of the splittings for
these $D$ and $B$ mesons.

One interesting phenomenon in QCD that we did not observe 
in our AdS/CFT model
is hyperfine splitting.
There are special pairs of mesons in QCD, which differ by the spin of
the heavy quark and for which
the mass difference is proportional to $\Lambda_{\rm{QCD}}^2 / m_h$.
In our fluctuation analysis, there are degeneracies in the spectra,
which
might provide a starting point to look for these hyperfine effects.
For example, the lowest lying excitation in the $x$ direction is a
vector meson with the
same energy as the scalar meson corresponding to the lowest lying
excitation in
the $y_6$ direction.  This degeneracy is likely a consequence of ${\mathcal N}=2$
supersymmetry, and we expect the fermionic fluctuations of the
superstring will fill out this ${\mathcal N}=2$ massive supermultiplet.
It is tempting to speculate that in a background with ${\mathcal N}=1$ or no
supersymmetry, the energies of the vector and scalar mesons will develop
a hyperfine splitting.\footnote{We would like to thank J.~Erdmenger
and D.~Son for discussion on this point.}

Finally, we make some comments regarding two specific open questions related to our work.

\subsubsection*{Hybrid mesons}

In phenomenological QCD literature, one finds discussion of hybrid mesons.
In perturbative language, such an object would be a bound state of a quark, antiquark,
and gluon \cite{jaffejohnson},
while at strong coupling, there exist models of a heavy quark and antiquark joined by
a vibrating flux tube \cite{isgurpaton}.  This second picture is similar to but also rather
different from our model.  Like us, the authors of Ref.~\cite{isgurpaton} begin by finding the modes
of the vibrating flux tube joining the quarks.  However, in their model, both quarks are heavy.
Also, and perhaps more importantly, the quarks themselves
have a mass large compared to the energy of the flux tube, whereas in ours, the mass of the
meson is the mass of the flux tube.  As a next step, the authors
of Ref.~\cite{isgurpaton} use the vibrating flux tube to construct a phenomenological
Cornell like potential through which the massive quarks interact.
Despite these differences, one wonders if there exists a closer connection between
our heavy-light mesons in ${\mathcal N}=2$ SYM and hybrid heavy-light mesons in QCD --- if such
things exist --- rather than the ``ordinary'' heavy-light mesons of QCD.

\subsubsection*{W bosons}

One may also consider Higgsing the ${\mathcal N}=4$ $\SUN$ SYM theory down to
$\mbox{SU}(N-2) \times U(1)^2$.  In the dual gravitational picture,
this Higgsing corresponds to pulling two D3 branes off of the stack of $N$ D3 branes,
whose low energy description this SYM theory is.  As long as we keep the D3 branes parallel in this $AdS_5 \times S^5$
geometry,
they do not experience a potential and we can imagine placing them
at nonzero values of $y$, just as we did for the D7 branes.  There is then a semi-classical
string that stretches between the two D3 branes, whose fluctuations we may study
and which has a dual field theory interpretation as a W boson.\footnote{%
 We would like to thank I.~Klebanov for suggesting we think about this extension of our results.
}

We mention this D3 brane and string construction because we can at this point
in our analysis treat it very easily.
The treatment of the string spinning in real space and corresponding to a heavy-light meson
is identical for the W bosons.  Also,
the $x$ fluctuations of such a string are identical
to the $x$ fluctuations for the heavy-light meson.  Finally, the $y_6$ fluctuations are
identical, except that there are now four additional $y_6$-like directions perpendicular
to the D3 brane string configuration.  Whereas for the heavy-light meson,
the $x$ and $y_6$ fluctuations gave us four towers of identical modes, and the
$\rho$ fluctuations gave us another four towers, for the W boson, the $x$ and $y_6$
fluctuations give us eight towers of identical modes.  We believe this regrouping
of one pair of four identical towers into eight identical towers is related to the doubling
in the amount of supersymmetry.  The ${\mathcal N}=2$ SYM relevant for the
heavy-light mesons has eight supercharges, whereas the ${\mathcal N}=4$ SYM,
after the Higgsing which breaks conformal invariance, should have 16.

\section*{Acknowledgments}
We would like to thank J.~Erdmenger, N.~Evans, Ph.~de Forcrand, K.~Kajantie,
D.~Kaplan, I.~Klebanov, A.~Kurkela, D.~Mateos, A.~Rebhan, D.~Rodriguez-Gomez
A.~Scardicchio, and D.~Son for valuable discussions.
We would also like to thank A.~Paredes and P.~Talavera for bringing their work 
\cite{ParedesTalavera} to our
attention.
C.P.H.\ would like to thank the KITP,
where part of this work was done, for hospitality.
C.P.H.\ was supported in part by the National Science Foundation
under Grants No.\  PHY-0243680 and PHY05-51164, S.A.S.\ by the Austrian Science Foundation, FWF, project No.~P19958, and
A.V.\ in part by the Austrian Science Foundation, FWF, project No.~M1006.



\end{document}